\documentstyle[preprint,aps]{revtex}
\begin{document}
\draft
\preprint{ 2DWC-Sliding-\today}
\title{
Sliding motion of a two-dimensional Wigner \\
crystal in a strong magnetic field }
\author{ Xuejun Zhu, P. B. Littlewood, and A. J. Millis}
\address{
AT\&T Bell Laboratories, Murray Hill, New Jersey 07974
}
\date{\today}
\maketitle

\begin{abstract}
We study the sliding state of a two-dimensional Wigner
crystal in a strong magnetic field and a random
impurity potential.
Using a high-velocity perturbation theory, we
compute the nonlinear conductivity, various
correlation functions, and
the interference effects
arising in combined AC + DC electric effects,
including the Shapiro anomaly and the linear response to an
AC field.
Disorder is found to
induce mainly transverse distortions in the sliding state of the
lattice.
The Hall
resistivity retains its classical value.
We find that, within the large velocity
perturbation theory, free carriers which affect
the longitudinal phonon modes of the Wigner crystal do not
change the form of the nonlinear conductivity.
We compare the
present sliding Wigner crystal in a strong
magnetic field to the conventional sliding charge-density
wave systems.
Our result for the nonlinear
conductivity agrees well with the $I-V$
characteristics measured in some experiments at low temperatures or large
depinning fields,
for the insulating phases near
filling factor $\nu$ = 1/5.
We summarize the available experimental data,
and point out the differences among them.

\end{abstract}

\pacs{PACS: 73.40.Kp,~73.20.Mf,~71.45.Lr}

\newpage
\section{Introduction}

The $T=0$ phase diagram of a two-dimensional (2D) electron gas in a strong
magnetic field is very rich. Two of the possible competing phases are
the Wigner crystal phase, which is believed to occur at sufficiently strong
magnetic field or low density, and the fractionally quantized Hall effect
(FQHE)
liquid phase, which occurs if the field is not too strong and the filling
factor $\nu$ has an odd denominator \cite{FQHE0,THEBOOK}. The
two-dimensional electron gas has been studied experimentally in
modulation doped semiconductor
heterojunctions.
Currently, the FQHE state
is firmly established down to filling factor $\nu = 1/5$
\cite{Jiang1} for samples with $n$-type dopants and
down to $\nu = 1/3$ for samples with $p$-type dopants. For $\nu$ slightly
greater than 1/5, but less than 2/9, and for $\nu$ smaller than
$1/5$,
insulating phases have been observed in the cleanest
$n$-type samples currently available
\cite{Jiang1,I,II,III,Jiang}.
There is also increasing experimental evidence that similar reentrant
insulating phases appear in $p$-type samples for $\nu$ slightly greater
and for $\nu$ less than $1/3$ \cite{hole}.
Most of the experimental data on the
insulating phases in this
regime are interpreted as due to
the Wigner crystallization
\cite{Jiang1,I,II,III,Jiang,hole,ZhuLouie}.

While the insulating nature of
the states around $\nu = 1/5$ has been clearly
demonstrated, it remains controversial if the insulating behavior
is primarily due to interaction-induced Wigner
crystallization or primarily due to disorder.
Nonlinearities in the conductivity have
been argued by several groups \cite{I,II,III,Jiang}
to imply that the
insulating phases are due to the formation of the Wigner crystal
which is then pinned by disorder.
However,
there exist some puzzling
discrepancies among experiments under
seemingly similar experimental conditions.
For example, the apparent depinning field (see Eq.~3)
differs by almost three
orders of magnitude between
two experiments
\cite{I,II} on samples with similar zero field mobility.
In another experiment \cite{Jiang}, nonlinearity in the
conductivity was observed in both of
these field ranges; but the
nonlinearity at larger fields
was attributed to electron heating effects at
larger currents.
Even within the small field ($\sim$
0.2~mV across samples with typical size $\sim$ $3~mm$)
nonlinear conductivity studies, some
significant differences exist among
experiments \cite{I,III,Jiang}.
We shall discuss these and other related experiments in more detail
later in the paper.

In view of the experimental situation, it is desirable
to provide a theoretical framework
for describing the properties of
a sliding
Wigner crystal in a strong magnetic field \cite{ZLMPRL}.
The principle theoretical issue is the proper treatment of the disorder which
distorts the Wigner crystal during sliding.
In this paper, we apply techniques originally developed for studying
sliding charge-density wave (CDW)
systems
\cite{CDWRev,Sneddon,MT,SNC,Vlattice}
to the present system.
We use
an elastic model for pinning
and for the interaction of the sliding Wigner crystal
with disorder, similar
to that used in the Fukuyama-Lee-Rice (FLR) model
for CDW's \cite{FLRMod},
and the high-velocity
perturbation theory \cite{Sneddon,MT}
to calculate various properties of a sliding Wigner crystal
in the presence of a strong magnetic field and disorder
potential.
This approach is expected to be valid
as
long as the sliding velocity is large, in comparison to
disorder effects, but not so large as to
invalidate the elastic medium theory.
In the context of
CDW systems,
similar perturbation theory \cite{Sneddon,MT}
was compared to numerical simulations, and was
found to be quantitatively
reliable for fields greater than $\sim$ 2 times the
threshold field.

The physics of a
Wigner crystal in a strong magnetic field differs that of a
conventional CDW because:
1) the presence of a
large magnetic field gives rise
to a large Hall component of the
motion; 2) the motion is two-dimensional whereas
it is one-dimensional in a conventional CDW system in arbitrary spatial
dimensions,
and the related issue of the shear modulus of
a Wigner crystal being much smaller
than the bulk modulus; 3) the presence of a weakly screened,
two-dimensional long-range
Coulomb interaction; 4) the order-parameter, {\it i.e.\/}, the
Fourier-transformed charge density, is non-zero on a two
dimensional grid, and does not decay very fast with increasing
reciprocal lattice vector $\vec G$.
The first point does not affect the
static properties of the Wigner crystal.
Points 2 and 3 have been shown to affect the
static properties, and together with point 1, the
$k$-dependence of the collective
excitations of a Wigner crystal in a significant way
\cite{NLM}. In the sliding regime where we are
interested in the dynamic response
of the Wigner crystal, they
will all affect the results.

Some recent experiments done at relatively high
temperatures \cite{Jiang} have also indicated the
possible involvement of the thermally excited free carriers in
the sliding Wigner crystal.
For example, the differential conductivity
continues to be thermally activated above the sliding threshold
in Ref.~\cite{Jiang}, up to a second threshold field,
thus raising the possibility that the thermally excited free
carriers play an important role in determining the
conductivity of the sliding state.
We have therefore considered,
within the same perturbation theory framework,
the effect of coupling to a
sliding Wigner crystal a set of free carriers which are
phenomenologically characterized by a conductivity tensor.
We find that the present perturbation
theory is unable to account for the experimental
results in Ref.~\cite{Jiang} between the two threshold
fields.
This suggests that the starting point of our theory,
{\it i.e.}, the assumption that we can
perturb around a uniformly sliding state, is probably not valid
in this regime.

We have also considered the effects of combined AC + DC fields,
where interference between the
internal ``washboard" frequency of the sliding
motion and the AC driving frequency can be used to further
identify the nature of the sliding state of the
insulating phases observed experimentally. These include the
Shapiro anomaly in the DC conductivity in the presence of
an AC field, and the AC response in the presence of a large
DC current. These effects are also studied using the
same perturbation theory, carried out to the same
order in disorder potential
as that for the DC nonlinear conductivity.

It has also been claimed recently that a pinned Wigner crystal
in zero external magnetic field exists at
$Si/SiO_2$ interfaces \cite{WCSi}. For completeness and
convenience of reference, we also give our results
for zero magnetic field, although they are
similar to those for the flux lattice sliding
motion in a type-II superconductor as studied by Schmid and
Hauger and by
Larkin and Ovchinnicov in Ref.~\cite{Vlattice}. The main difference
here is that
for the present Wigner crystal case the long-range Coulomb interaction
yields a $k^{1/2}$-dispersive longitudinal phonon mode, and this
dispersion could be affected by the presence of free-carriers
in ways that are described in \cite{NLM} and also discussed
later in this paper. As we
will show, however, the main distortions in a sliding
Wigner crystal are transverse; and many properties
of a sliding Wigner crystal are not affected by the longitudinal
mode spectrum.

The balance of the paper is as
follows. In Sec.~II, we introduce our model,
make some general remarks regarding the Hall
resistivity, and summarize recent experiments concerning the Hall
resistivity for the insulating phases around $\nu = 1/5$.
In Sec.~III, we
study
the nonlinear conductivity of the Wigner crystal in the sliding
regime in the absence of free carriers, treating
the scattering by impurities as a
perturbation. Various correlation functions
of the sliding motion are studied in Sec.~IV.
The effects of the disparity of the
bulk modulus and the shear modulus are
illustrated.
In Sec.~V, we consider the
effects of free carriers on the sliding dynamics of
the Wigner crystal.
In Sec.~VI, we consider the two AC + DC interference
effects mentioned above.
Results for the zero magnetic field case are given in
Sec.~VII.
All of our main results  for the transport properties
of a sliding Wigner crystal
are summarized in Sec.~VIII.
In Sec.~IX, we discuss in some detail the current experimental
situation and attempt to make contact
between our theory and these experiments.
Sec.~X is the conclusion. Readers who are only interested in the
results and the current experimental status may go to Secs.~VIII,
IX and X directly.

\section{Equation of Motion and the Hall Resistivity}

In Sec.~IIA, we first establish the equation
of motion. We remark briefly on its
similarities to and differences from the
equations of motion of related models.
In Sec.~IIB, we give results for
the Hall resistivity
that can be obtained immediately by examination of the
equation of motion.
The steady state solution to the equation
of motion in the absence of random potential scattering
is discussed in Sec.~IIC.

\subsection{Model and Equation of Motion}

We begin by considering the various
forces acting on an element of the
Wigner crystal at $(\vec r, t)$ in two dimensions,
which we treat classically as an elastic medium.
We denote the two-dimensional displacement of the Wigner crystal at
$(\vec r, t)$ as $\vec u(\vec r, t)$,
average mass and charge densities as $\rho_m$ and $\rho_c$,
lattice constant of the Wigner crystal as $a$, and
background dielectric constant $\epsilon$.
The following terms enter the equation of motion:

1) inertial force: $\rho_m$ $\ddot {\vec u}$;

2) dissipation: $\lambda \rho_m \dot {\vec u}$.
$\lambda$ is a phenomenological
constant describing damping due to
everything other than the
disorder potential that is treated
explicitly below;

3) Lorentz force:
$\rho_m \omega_c \vec e_z  \times \dot {\vec u}$,
where $\omega_c = \rho_c B/\rho_m$ is the cyclotron frequency,
$\vec e_z$ is the unit vector in the direction
of the perpenticular magnetic field;

4) elastic restoring force:
$\int {\bf D}(\vec r- \vec {r'}) \cdot ({{\vec u}(\vec {r'}, t) - \vec v t})
d^2\vec {r'}$. Here ${\bf D}$ is the real-space
dynamic matrix tensor, and $\vec v t$ is the uniform component
of $\vec u(\vec r,t)$ in the sliding state of the Wigner crystal defined
as $\vec v t = {lim}_{T\rightarrow\infty}
               {lim}_{V\rightarrow\infty} \int{dt \over T}
               \int{{d\vec r} \over V} \vec u (\vec r,t)$.
Upon Fourier transformation, the
component transverse to $\vec k$ is
${\bf D}^T(\vec k)/\rho_m = c^T k^2$ and the
longitudinal component ${\bf D}^L(\vec k)/\rho_m = c^L k$.
These are valid in the long
wavelength limit, {\it i.e.\/}, for
$|\vec u(\vec r) - \vec u(\vec {r'})|$
much less than $|\vec r - \vec {r'}|$.
Following Ref.~\cite{NLM},
we have: $c^T = \Omega^2a^2\alpha$,
and the $k$-dependent
$c^L = \Omega^2(2\pi a + (1+\alpha) a^2 k)$,
where
$\alpha \sim 0.02$ is the ratio
of the shear modulus with the bulk modulus, and
$\Omega^2 = \rho_c^2 / \rho_m a\epsilon$ is a characteristic
frequency of the Wigner crystal, on
the order of the zone-boundary
phonon frequency at zero magnetic field.
The first term in $c^L$ is due to
the long-range
Coulomb interaction.
Here we view
the Wigner crystal
as a deformable classical solid characterized by
its dynamic matrix ${\bf D}(\vec k)$, given by, for example,
Bonsall and Maradudin
\cite{BM77}. Quantum mechanical corrections
have been studied within the time-dependent Hartree-Fock
theory and are found to be on the order of $10\%$ for the
long wavelength phonons that we are interested in
\cite{Cote};

5) disorder potential which
couples to the $\vec r$-dependent charge
density
$\rho(\vec r)$:
$\rho(\vec r) \nabla_r V_{dis}(\vec r + \vec u(\vec r,t)).$ Here
$V_{dis}(\vec r)$ is the disorder potential at $\vec r$. Only
the correlation function of $V_{dis}$ will
enter our results, so we will not specify its
form. In cases where $V_{dis}$ is due to
the remote dopants at a distance $d$ away from the plane
of the 2D electron gas, the Fourier
transformation of $V_{dis}(\vec r)$, $V_{dis}(\vec q)$
decays as $e^{-qd}$ from simple eletrostatic considerations;

6) and finally, the external
driving force on the
total charge density $\rho_c \vec E^{ext}$.

Combining the six terms, we obtain the
equation of motion in real
space:

\begin{equation}
{\rho_m \ddot {\vec u} +
\lambda \rho_m \dot {\vec u} +
\rho_m \omega_c \vec e_z \times \dot {\vec u} +
\int {\bf D}(\vec r- \vec {r'}) \cdot \vec u(\vec {r'}, t) d^2\vec {r'} =
\rho_c \vec E^{ext} -
\rho(\vec r) \nabla_{\vec r} V_{dis}(\vec r + \vec u(\vec r,t))}.
\label{EOM}
\end{equation}

In a perfectly ordered Wigner crystal,
the charge density is determined by its Fourier
components $\rho (\vec G)$ at
the reciprocal lattice vectors $\vec G$, via,
$\rho (\vec r) = \sum_{\vec G}
\rho (\vec G)e^{i\vec G \cdot \vec r}$.
In the systems of interest, the magnetic field is so strong
that the magnetic length
$l_B = \sqrt{\hbar c/eH}$ is smaller than the lattice constant, and
$\rho (\vec G)$
decays
as $\rho (\vec G) = e^{-G^2 l_B^2/2}$,
so many reciprocal lattice vectors will be important,
in contrast to the CDW case in which only the
lowest $\vec G$ is important.

For later reference, we write down here also the
equation of motion in the FLR model for a CDW. Apart from the
absence of a magnetic field, the most important
difference is that because of the
quasi-one dimensional crystal structure of CDW systems,
the electrons are free to move only in one direction, which we take to be
$z$. Therefore:
\begin{equation}
{\rho_m \ddot {u_z} +
\lambda \rho_m \dot {u_z} +
\int {\bf D}_{zz}(\vec r- \vec {r'}) u_z(\vec {r'}, t) d\vec {r'} =
\rho_c E_z^{ext} -
\rho(\vec r)
{\partial \over {\partial z}}V_{dis}(\vec r + \vec {e_z} u_z(\vec r,t))}.
\label{EOMFLR}
\end{equation}
Dimensionality, which is crucial in determining
the functional form of the nonlinear
conductivity of a sliding CDW \cite{Sneddon},
enters through the elastic term in the above
equation.

In the absence of the external
potential, very weak disorder
can pin the overall phase of the charge-density wave
\cite{FLRMod}.
In this case, the phase fluctuations diverge beyond
a length $\xi^T$ which defines a typical domain
size in a pinned CDW.
Similar considerations apply here to the Wigner crystal case
as well \cite{NLM}.
As a result of pinning, the linear response conductivity
vanishes.
But when $E^{ext}$ is greater than some
threshold field $E_{Th}$, the pinned Wigner crystal will
begin to slide. The pinning-depinning transition between the static
state and the sliding state in a CDW system has been argued to be
second order, exhibiting the critical phenomena \cite{DFisher}.

In the weak pinning limit, following the Fukuyama-Lee-Rice
argument \cite{FLRMod,NLM}, one can relate $E_{Th}$
to the static
correlation length $\xi^T$ of the pinned Wigner crystal as follows.
At the threshold
field $E_{Th}$, the total force
acting on a pinned domain of Wigner crystal of linear
size $\xi^T$ is $E_{Th}\rho_c\pi(\xi^T)^2$.
This should be balanced by the total elastic force
acting on the same domain. Since by definition, the
displacement over the length $\xi^T$ is
$\sim a$, on the order of the lattice constant,
then the total elastic restoring force must be
$\rho_m\Omega^2a^2\alpha \times a$. Therefore,
\begin{equation}
E_{Th}
\approx {{\rho_m\Omega^2a^2\alpha}\over {\rho_c (\xi^T)^2\pi/a}}.
\label{ETH}
\end{equation}
We can also define \cite{NLM} a longitudinal correlation length
$\xi^L$ beyond which the Coulomb interaction energy is
smaller than the pinning energy:
$c^L/\xi^L = c^T/(\xi^T)^2$,
or $\xi^L = 2\pi (\xi^T)^2 / (a\alpha)$.
Thus there are two length scales involved in the
present problem, related to the shear
modulus and the long range Coulomb interaction
respectively. In the case where the latter is screened
by free carriers, $\xi^L$ will be larger than $\xi^T$
by at least a factor of $1/\sqrt{\alpha}$.

Various aspects of the FLR model have been
studied
in the $B=0$ charge-density wave context
\cite{CDWRev,Sneddon,MT}.
The linear response of the pinned state for $B \ne 0$ \cite{NLM,FL}
has also been determined.
In Ref.~\cite{NLM}, the large difference
between the bulk modulus and
the shear modulus was taken into
account. One consequence is the
occurence of a $k^{1/2}$-dispersing
mode for $k$ lying between
$1/\xi^L$ and $1/\xi^T$
\cite{II,NLM}.
In the absence of disorder, or for $k$ greater than
$1/\xi^T$ in a statically disordered lattice,
there exists a $k^{3/2}$-dispersing mode
which
is characteristic of a 2D Wigner crystal in a strong magnetic
field; it results from the mixing due to the magnetic field
of the transverse and the longitudinal phonon modes which
disperse as $k$ and $k^{1/2}$ respectively when there is no magnetic
field, as one can see from the small $k$-behavior of the
dynamic matrix given earlier.
The present study focuses on the
sliding regime, treating the
disorder potential as a perturbation
relative to the external driving
force.

We note that although the present problem bears some
resemblance to the sliding motion of the vortex lattice in
a type-II superconductor \cite{Vlattice},
in that there is a Lorentz force,
and that the shear modulus is much smaller than the
bulk modulus, there is a very important difference.
The sliding motion of the vortex
lattice is driven by
the Lorentz
force due to the superconducting
current. The moving vortex lattice then in turn induces a
voltage drop in the direction
of the superconducting current.
In the present problem, the depinning electric force is supplied externally
and the magnetic field enters explicitly the
equation of motion.

The present model may be used to study the sliding
of a 2D Wigner crystal with no external magnetic field by simply
setting the Lorentz force to zero. Such a system has been shown
to exist for electrons on a helium film \cite{Jiang89}, and more recently,
has been argued
for doped electrons at $Si/SiO_2$ interfaces \cite{WCSi}.
We will also give our results for cases where
there is no external magnetic field. When expressed in terms
of the shear modulus, they are basically the same as those
for a sliding vortex lattice \cite{Vlattice}.
As we mentioned already, the difference in the longitudinal phonon spectrum
between the two systems
does not appear in the large velocity limit.

\subsection{The Hall Resistivity }

Before we embark upon a detailed
study of Eq.~\ref{EOM}, it is worth
noting that one can obtain the Hall {\it resistivity}
$\rho_{xy}$ immediately
from Eq.~\ref{EOM}
for a Wigner crystal in the sliding state.
With a constant current density
$j_x$ in the $x$-direction, the Hall resistivity
relates $E^{ext}_y$ to $j_x$ via $E^{ext}_y = \rho_{yx} j_x$.
We disorder-average the Fourier transformed
Eq.~\ref{EOM} under the steady state
condition $\dot {u_x} \not= 0$.
Assuming isotropy, we obtain (using $j_x = \rho_c \dot {u_x}$):
\begin{equation}
\rho_c E^{ext}_y = \rho_m \omega_c \dot {u_x}
=\rho_c\biggl( {B \over \rho_c} \biggr) j_x,
\label{Hall}
\end{equation}
so $\rho_{yx} = B / \rho_c e$.
This argument is independent
of the form of the disorder potential (aside from isotropy),
and requires only that
the current be transported by a $k = 0$ mode of the crystal.
We note however that these arguments are strictly only valid
at $T = 0$, where there are no thermally
activated excitations that carry current.
Effects of nonzero
temperatures on the Wigner crystal
state below the sliding threshold
are not well understood, and in general
will depend on the nature of the possible
charged thermal excitations in the system.

The issue of Hall {\it resistivity} for the
insulating phases in the FQHE regime
was raised in how to best characterize
the nature of the insulating phase
\cite{CSGL}. Kivelson,
Zhang, and Lee \cite{CSGL} have suggested
that it be better described as a
``Hall insulator", which they characterize
by a diverging longitudinal
resistivity and a constant
Hall resistivity at zero temperature.
A more concrete description of the
``Hall insulator" in the FQHE regime has not been given.
Several authors have recently argued that
an Anderson insulator is in fact also a ``Hall
insulator" with Hall resistivity given
approximately by $B/\rho_ce$, $ \rho_c$ being the total localized
carrier density \cite{CSGL,Imry}.

Experimental results for $\rho_{xy}$ in the insulating
phases have now become available \cite{expt}.
They are obtained
by measuring the differential Hall
resistance at a constant current.
The mixture of $\rho_{xx}$ into $\rho_{xy}$, which can be
rather large at low temperatures, can be subtracted
by an average
over $\rho_{xy}(\vec B)$ and $\rho_{xy}(- \vec B)$
\cite{expt}. The experimental findings can
be summarized as follows:

\noindent
1). At low temperatures and well below the sliding
threshold, the Hall resistivity is given by
$\rho_{yx} \stackrel{\sim}{=} B/\rho_c$,
where
$\rho_c$ is the total electron density that was introduced into
the sample during doping, and therefore $\rho_{xy}$ is
insensitive to temperature,
in sharp contrast to $\rho_{xx}$
in the same regime which is often thermally activated. When
it was measured, the low-frequency response was
always found to be resistive, with very small out-of-phase
component in these systems
\cite{expt}.

\noindent
2). The Hall resistivity is found to have little temperature
or field
dependence above and below the so-called critical
temperature which is marked by the absence of an apparent
depinning transition. The longitudinal resistivity changes behavior:
it may or may not show thermally-activated temperature
dependence; also it may or may not be ohmic \cite{expt}.

If the applied field is above the
sliding threshold for the pinned Wigner crystal, our argument above
can be used to argue for a normal Hall resistivity.
It is not clear at this point
if the experimental data below the apparent sliding
threshold could be understood within the Wigner crystal picture.

It is perhaps worth commenting that at $T = 0$ the Hall
resistivity of a {\it pinned} Wigner crystal within the linear
response regime, {\it i.e.}, below sliding threshold can
also be shown to be simply $B/\rho_c$, following earlier
work by Fukuyama and Lee on the magneto-conductivity
of a pinned CDW \cite{FL}.
This conclusion can be generalized at a finite frequency
below sliding threshold,
by repeating the argument
given at the beginning of this subsection leading to Eq.~\ref{Hall},
thus avoiding the
need to resort to the particular form of the pinned phonon spectrum
as was done in \cite{FL}.
These arguments, as well as those
for Anderson insulators \cite{CSGL},
are made at $T=0$ where there are no
thermal excitations in the insulator.
However, this mechanism necessarily requires that
the conduction be carried by the
polarization current excited at a finite frequency, implying
a capacitive response. It then
follows that the measured current and the voltage must be
out of phase by $90^o$. This scenario is disfavored by
the current experiments \cite{expt}.
More careful measurements geared specifically toward this issue
would certainly be very useful, however.

A final comment is in order about the distinction between resistivity and
conductivity in a disordered system, when the
response functions become non-local and local field effects
are important. In particular, the
conductivity tensor is defined by $\vec j(\vec q) = \sum_{\vec {q'}}
\stackrel{\leftrightarrow}{\sigma} (\vec q,\vec {q'}) \cdot \vec E(\vec {q'})$;
$\sigma$ becomes diagonal only after averaging over disorder. The resistivity
is the inverse of the conductivity matrix in $\vec q$-space. The
general arguments we have given in this section relate to
$\rho_{xy}$ defined as
$lim_{\vec q,\vec {q'} \rightarrow 0} \sigma_{xy}^{-1}(\vec q,\vec {q'})
\neq
[lim_{\vec q,\vec {q'} \rightarrow 0} \sigma_{xy}(\vec q,\vec {q'})]^{-1}$.
This is the resistivity measured when the current flows entirely
in the uniform mode, and may not correspond to the
simple inversion of a measurement of the conductance.

\subsection{Steady State Solution with No Disorder}

For $\nabla_{\vec r}V_{dis} =0$,
the equation of motion
Eq.~\ref{EOM},
can be solved trivially. In steady state:
\begin{equation}
{\rho_c \over \rho_m} \vec E^{ext} =
\left(\matrix{\lambda&-\omega_c \cr
  \omega_c&\lambda\cr}\right)
\cdot \vec v, ~{\rm and}~, \vec u(\vec r,t) = \vec v t,
\label{NODis}
\end{equation}
with
$v = {\rho_c \over {\rho_m\sqrt{\lambda^2+\omega_c^2}}}
E^{ext}$.
The resistivity tensor ${\bf \rho}$ is simply:
\begin{equation}
{\bf \rho} =
{\rho_m \over \rho_c^2}
\left(\matrix{\lambda&-\omega_c \cr
  \omega_c&\lambda\cr}\right),
\label{NODisrho}
\end{equation}
and the conductivity tensor $\bf \sigma$ is:
\begin{equation}
{\bf \sigma} =
{\rho_c^2 \over {\rho_m{(\lambda^2+\omega_c^2)}}}
\left(\matrix{\lambda& \omega_c \cr
  -\omega_c&\lambda\cr}\right).
\label{NODissig}
\end{equation}

If there is an additional physical dissipative process
which changes
$\lambda$ by a small amount $\delta \lambda$, we will have
for the components of the resistivity tensor
\begin{equation}
{{\delta \rho_{xx}} \over {\delta \lambda}}
= {{\rho_m} \over {\rho_c^2}},
\label{changerho1}
\end{equation}
and,
\begin{equation}
{{\delta \rho_{xy}} \over {\delta \lambda}}
= 0;
\label{changerho2}
\end{equation}
and for the conductivity tensor
\begin{equation}
{{\delta \sigma_{xx}} \over {\delta \lambda}}
= {{\rho_c^2} \over {\rho_m}}
{{\omega^2_c - \lambda^2} \over {(\omega^2_c + \lambda^2)^2}},
\label{changesig1}
\end{equation}
and,
\begin{equation}
{{\delta \sigma_{xy}} \over {\delta \lambda}}
= {{\rho_c^2} \over {\rho_m}}
{{-2\omega_c\lambda} \over {(\omega^2_c + \lambda^2)^2}}.
\label{changesig2}
\end{equation}

One sees that for $\lambda < \omega_c$, $\sigma_{xx}$ increases with
$\lambda$, in contrast to the zero-magnetic-field case.
In addition,
$|\delta\sigma_{xx}| \gg |\delta\sigma_{xy}|$ while
$|\sigma_{xx}| \ll |\sigma_{xy}|$ for $\lambda \ll \omega_c$.

And finally,
\begin{equation}
{\delta v \over v} =
{{-\lambda \delta\lambda} \over {\lambda^2+\omega^2_c}}.
\label{changeV}
\end{equation}

If one views the disorder scattering as an additional
source of dissipation which renormalizes $\lambda$,
many of these observations can be carried over. The problem
therefore is to find how disorder scattering modifies
$\lambda$. To this end, a perturbative analysis is carried
out in Sec.~III.

\section{Perturbation Theory: Nonlinear Conductivity}

We give here the detailed results for the nonlinear
conductivity of a sliding Wigner crystal in the presence of a
strong perpendicular magnetic field with disorder
scattering.
Results here are also used in the next
section for studying the correlation functions of the sliding
Wigner crystal.
The dependence of the Hall angle on the sliding current
is calculated. In Sec.~IIIA, we introduce the response
function for the present system, and we outline the
perturbation theory we use.
In Sec.~IIIB, we solve for the velocity to the second order
in disorder scattering, thereby obtaining the nonlinear conductivity.
We show in Sec.~IIIC that the large-velocity perturbation theory
also gives a measure of the depinning field.
The Hall angle is studied in Sec.~IIID.

\subsection{The Perturbation Series and the Green's Function}

We shall in this section restrict ourselves to a DC electric field
where most of the experiments are carried out.
The extension to the more general DC + AC field case
will be discussed in Sec.~VI.
It is convenient in the present problem to
write the DC external field in terms of a velocity
$\vec v_0$ as:
\begin{equation}
{\rho_c \over \rho_m} \vec E^{ext} =
\left(\matrix{\lambda&-\omega_c \cr
  \omega_c&\lambda\cr}\right)
\cdot \vec v_0,
\label{field}
\end{equation}
and the displacement as:
\begin{equation}
\vec u(\vec r,t) =
\vec v_0 t + \vec x(\vec r,t),
\label{dis}
\end{equation}
where the first term would be the correct solution to
Eq.~\ref{EOM} for the field in Eq.~\ref{field}
if there was
no disorder (see Eqs. \ref{NODisrho}~and~\ref{NODissig}).
The last term, $\vec x$, represents the
additional displacement due to disorder scattering.
By substituting Eq.~\ref{dis} into
Eq.~\ref{EOM} and using Eq.~\ref{field}, we find that
$\vec x(\vec r,t)$ satisfies the following equation of motion:
\begin{equation}
{ \rho_m \ddot {\vec x} +
 \rho_m \left(\matrix{\lambda & -\omega_c \cr
                      \omega_c  & \lambda\cr} \right) \cdot  \dot{\vec x} +
\int {\bf D}(\vec r- \vec {r'}) \cdot \vec x(\vec {r'}, t) d\vec {r'} =
-\rho(\vec r) \nabla_{\vec r} V_{dis}(\vec r + \vec u(\vec r,t)).}
\label{EOM1}
\end{equation}

Formally, the solution of the above
equation can be written in terms of
tensor Green's function  ${\bf G}$
as:
\begin{equation}
\vec x(\vec r,t) = \int d\vec {r'} dt' {\bf G}(\vec r - \vec {r'}, t-t') \cdot
\biggl[ - \rho (\vec {r'})
\nabla_{\vec {r'}} V_{dis}(\vec {r'} + \vec u(\vec {r'},t'))
\biggr],
\label{xandg}
\end{equation}
where ${\bf G}$ is the solution of:
\begin{equation}
{\rho_m \ddot  {\bf G}(\vec r,t)  +
\rho_m \left(\matrix{\lambda & -\omega_c \cr
\omega_c  & \lambda\cr} \right) \cdot {\dot {\bf G}(\vec r,t)} +
\int {\bf D}(\vec r- \vec {r'}) \cdot  {\bf G}(\vec {r'}, t) d\vec {r'} =
\delta (t) \delta (\vec r) \left(\matrix{1 & 0 \cr 0 & 1 \cr} \right).}
\label{G}
\end{equation}

Fourier transforming Eq.~\ref{G}, using
the long-wavelength limit expression for
${\bf D}$ which we gave in Sec.~IIA, and assuming
$\omega \ll \lambda$, we have:
\begin{equation}
{\bf G}^{-1} (\vec k, \omega)/\rho_m = c^Lk     {{\vec k\vec k}\over k^2}
                         +c^Tk^2\bigl(1-{{\vec k\vec k}\over k^2}\bigr)
                         - i\omega
\left(\matrix{\lambda & -\omega_c \cr \omega_c & \lambda \cr}\right),
\label{Gk1}
\end{equation}
the inverse of which can be taken easily:
\begin{equation}
{\bf G} (\vec k, \omega)
=
{1 \over {P(\vec k,\omega)}}
 \Biggl[ c^Lk (1-{{\vec k\vec k}\over k^2})
                         +c^Tk^2{{\vec k\vec k}\over k^2}
                         - i\omega
\left(\matrix{\lambda & \omega_c \cr - \omega_c & \lambda \cr}\right)
  \Biggr].
\label{Gk2}
\end{equation}
The poles in ${\bf G}$ are given by the zeroes of $P(\vec k,\omega)$
which is:
\begin{equation}
P(\vec k,\omega) = 2\pi\Omega^4\alpha \rho_m
\biggl[(ka)^3 - \kappa_{\omega}^3 -
i(ka) \kappa_{\lambda}^2 sgn(\omega)\biggr].
\label{gpoles}
\end{equation}
In the above equation, we have defined two dimensionless
quantities:
\begin{eqnarray}
\kappa^3_{\omega}
&=& {{\omega^2(\omega^2_c+\lambda^2)} \over {2\pi\Omega^4\alpha}},
\nonumber \\
\kappa^2_{\lambda}&=& {{\omega\lambda} \over {\Omega^2\alpha}}.
\label{kdefs}
\end{eqnarray}
We have also
ignored the $\alpha (ka)^2$ term
whenever it appears together with $(ka)$ term since
$\alpha$ is small and the relevant $(ka)$'s are also small.
$\kappa_{\omega}/a$ is the wavevector at which the $k^{3/2}$-mode
is at frequency $\omega$, and $\kappa_{\lambda}$ measures the
strength of dissipation at the frequency $\omega$.
Their physical meaning for the
sliding state will become more clear later
in the paper.

Eq.~\ref{xandg} can be used to generate a perturbation series for
$\vec x(\vec r,t)$ in terms of the disorder
potential: to first order, $\vec x_1$ is given by Eq.~\ref{xandg}
with $\vec u(\vec {r'},t')$ in $V_{dis}$ on the right-hand-side
replaced
by $\vec v t'$; and the second correction $\vec x_2$ is given by
replacing $V_{dis}\bigl(\vec {r'} + \vec u(\vec {r'},t')\bigr)$ by
$\vec x_1(\vec {r'},t') \cdot
\nabla_{\vec {r'}}V_{dis}(\vec {r'}+\vec v t')$; {\it etc.}

\subsection{Nonlinear Conductivity}

Following the last subsection, we obtain for the displacement
to first order in the disorder potential $V_{dis}$:
\begin{equation}
\vec x(\vec r,t) = \int {{d^2\vec k} \over {(2\pi)^2}}
{{d\omega} \over {2\pi}} {{d^2\vec q} \over {(2\pi)^2}}
e^{ i \vec k \cdot \vec r - i \omega t}
\vec F_1(\vec k,\vec q,\omega),
\label{x1}
\end{equation}
where $\vec F_1(\vec k,\vec q,\omega)$ is given by
\begin{equation}
\vec F_1(\vec k,\vec q,\omega)
=-\rho(\vec k-\vec q) \biggl[{\bf G}(\vec k,\omega) \cdot(i\vec q)\biggr]
V_{dis}(\vec q)
2\pi
\delta (\omega + \vec q \cdot \vec v);
\label{F1}
\end{equation}
and for the velocity to second order in $V_{dis}$:
\begin{equation}
\rho_m \left(\matrix{ \lambda
& -\omega_c \cr \omega_c & \lambda\cr}\right)
\cdot \langle \dot {\vec x} \rangle
=
\sum_{\vec G} |\rho(\vec G)|^2
\int {{d^2\vec k} \over {(2\pi)^2}} ~~\vec F_2(\vec k, \vec G),
\label{v2}
\end{equation}
where $\vec F_2(\vec k, \vec G)$ is:
\begin{equation}
\vec F_2(\vec k)= \vec k
\Gamma(\vec k)
\biggl[ \vec k \cdot i{\bf G}(-\vec k + \vec G,
\vec k \cdot \vec v)
\cdot \vec k \biggr].
\label{F2}
\end{equation}

In Eq.~\ref{v2}, we
have averaged over disorder and integrated over the
$\delta$-function.
$V_{dis}(\vec q) = \int d^2\vec r e^{-i\vec q \cdot
\vec r} V_{dis}(\vec r)$,
and $\Gamma(\vec k)$ is the disorder correlation function:
$<V_{dis}(\vec k)V_{dis}(\vec q)>
= \delta (\vec k + \vec q) \Gamma (\vec k)$, where $<...>$ indicates
averaging over disorder.
We have also used the fact that for an undistorted Wigner cyrstal:
\begin{equation}
\rho(\vec q) = (2\pi)^2 \sum_{\vec G} \rho(\vec G) \delta(\vec q - \vec G),
\label{rhoG}
\end{equation}
where $\vec G$'s are reciprocal lattice vectors.

Using Eqs.~\ref{F1} and \ref{F2} in Eqs.~\ref{x1} and \ref{v2},
we have:
\begin{equation}
\vec x(\vec r,t) = \int {{d^2\vec k} \over {(2\pi)^2}}
{{d^2\vec q} \over {(2\pi)^2}}
e^{ i \vec k \cdot \vec r + i (\vec q \cdot \vec v) t}
\biggl[-\rho(\vec k-\vec q)\biggr]
\biggl[{\bf G}(\vec k, -\vec q \cdot \vec v) \cdot(i\vec q)\biggr]
V_{dis}(\vec q),
\label{x11}
\end{equation}
and,
\begin{equation}
\rho_m \left(\matrix{ \lambda
& -\omega_c \cr \omega_c & \lambda\cr}\right)
\cdot \langle \dot {\vec x} \rangle
=  \sum_{\vec G} |\rho(\vec G)|^2
\int {{d^2\vec q} \over {(2\pi)^2}}~~ \vec q \Gamma(\vec q)
\biggl[ \vec q \cdot (-Im {\bf G}(-\vec q + \vec G,\vec q \cdot \vec v))
\cdot \vec q \biggr].
\label{v22}
\end{equation}

In both Eq.~\ref{x11} and Eq.~\ref{v22},
the intergration over $\vec k$ is over the
first Brillouin zone, and that over $\vec q$ is over the whole
reciprocal lattice space.
Eq.~\ref{x11} is used later for
studying various correlation functions in
a sliding Wigner crystal.
We shall now focus on
Eq.~\ref{v22} which gives the lowest
order correction to the
conductivity. It is most easily studied using the
representation of ${\bf G}$ in Eq.~\ref{Gk2}.

Before we proceed further, we should mention that an alternative view
of the same perturbation theory is to ask: In order to
sustain the current flow $\rho_c \vec v$ in
the presence of $V_{dis}$, what additional
external field one would need above that given by Eq.~\ref{field}.
The answer is that to second order in disorder potential
the additional field $\vec {E^p}$
is given by:
\begin{equation}
\rho_c \vec {E^p}
=  \sum_{\vec G} |\rho(\vec G)|^2
\int {{d^2\vec q} \over {(2\pi)^2}}~~ \vec q \Gamma(\vec q)
\biggl[ \vec q \cdot Im {\bf G}(-\vec q + \vec G,\vec q \cdot \vec v)
\cdot \vec q \biggr].
\label{addE}
\end{equation}

The integral in Eq.~\ref{v22} or
Eq.~\ref{addE} is dominated
by two infared terms where the
elastic medium theory is expected to be valid.
They are respectively related to $\kappa_{\omega}$
and $\kappa_{\lambda}$ in Eq.~\ref{gpoles}.
For the convenience of the reader, we rewrite them here at
$\omega = \omega_v = |\vec G\cdot\vec v|$:
\begin{eqnarray}
\kappa^3_{\omega} &=&
{{\omega_v^2(\omega^2_c+\lambda^2)} \over {2\pi\Omega^4\alpha}},
\nonumber \\
\kappa^2_{\lambda}& =& {{\omega_v\lambda} \over {\Omega^2\alpha}}.
\label{krraydefs1}
\end{eqnarray}
We have suppressed the dependence of
$\omega_v$, $\kappa_{\omega}$ and
$\kappa_{\lambda}$ on $\vec G$
for ease of writing. Both $\kappa_{\omega}$ and $\kappa_{\lambda}$
must be much less than unity in order for the
elastic medium theory to be valid, and
their ratio
$\kappa_{\omega}/\kappa_{\lambda}
\sim (\omega_v\omega_c^4\alpha / 4\pi^2\lambda^3\Omega^2)^{1/6}$
depends on the sliding velocity $\vec v$ through $\omega_v$.
The dependence is not very strong, $\sim~ v^{1/6}$.

The result of the integral in
Eq.~\ref{v22} depends on the relative
size of $\kappa_{\omega}$ versus $\kappa_{\lambda}$.
We find that in general:
\begin{eqnarray}
\left(\matrix{ \lambda & -\omega_c \cr \omega_c & \lambda\cr}\right)
\cdot \langle \dot {\vec x} \rangle
&=& - \alpha_0 {1 \over \rho_m}
\sum_{\vec G}
A(\vec G)
\vec G ~sgn({\vec G \cdot \vec v}) \nonumber \\
&=& -\alpha_{dis}{{\vec v} \over v},
\label{v221}
\end{eqnarray}
where
\begin{equation}
A(\vec G) = {1 \over 4\pi\rho_mc^T} |\rho(\vec G)|^2
\Gamma(\vec G) |\vec G|^2.
\label{defAG}
\end{equation}
The constant $\alpha_0$ in Eq.~\ref{v221}
is $\pi/3$ if $\kappa_{\omega} \gg \kappa_{\lambda}$ and
is $\pi/4$ if $\kappa_{\omega} \ll \kappa_{\lambda}$, and changes
monotonically between these limits.
$\alpha_{dis}$ is uncertain
up to a numerical prefactor of order unity
that depends on the direction of $\vec v$
relative to the lattice orientation.
The relative stability of various sliding directions
was discussed in
\cite{Vlattice} in the sliding vortex lattice
context from a minimum-entropy-generation argument.
In this work, we simply assume that the Wigner crystal
is sliding along one of the stable configurations which may be
defined with a vector $\vec G$.

Therefore, if we use the alternative approach given by Eq.~\ref{addE}, we find
that for a given average sliding velocity $\vec v$,
the total applied field $\vec E^{ext}$
must be:
\begin{equation}
\vec E^{ext} = {\rho_m \over \rho_c} \biggl[
\left(\matrix{ \lambda & -\omega_c \cr \omega_c & \lambda\cr}\right)
\cdot \vec v + \alpha_{dis}{\vec v \over v} \biggr].
\label{addEsol}
\end{equation}

It is important to note that the right-hand-side of
Eq.~\ref{v221} does not depend on the magnitude of
$\vec v$, only on its direction. This feature is common to
all the 2D systems studied within the second order
Born approximation, and has
been noted in the CDW context in,
{\it e.g.}, Ref.~\cite{Sneddon}.
Eq.~\ref{v221} also shows that the net effect of disorder
is to increase the $\lambda$ in the equation of motion
by an amount of $\alpha_{dis}/v$.

It is straightforward to obtain the total velocity
$\vec v + \langle  \dot {\vec x} \rangle$, for a  given external
field $\vec E^{ext}$ in Eq.~\ref{field}.
The resultant conductivity tensor
has components given by:
\begin{equation}
\sigma_{xx} = {\rho_c^2 \over \rho_m}
    { 1 \over {\lambda^2+\omega_c^2}}
\biggl[\lambda + {\alpha_{dis} \over v} {{\omega_c^2 - \lambda^2}
                                  \over {\omega_c^2 + \lambda^2}}
\biggr],
\label{sigmaxx}
\end{equation}
and
\begin{equation}
\sigma_{xy} = - \sigma_{yx} = {\rho_c^2 \over \rho_m}
    { 1 \over {\lambda^2+\omega_c^2}}
\biggl[ \omega_c - {\alpha_{dis} \over v} {{2\omega_c\lambda}
                                  \over {\omega_c^2 + \lambda^2}}
\biggr],
\label{sigmaxy}
\end{equation}
Or equivalently in terms of the resistivity to order
$\alpha_{dis}$:
\begin{equation}
\rho_{xx} = {\lambda\rho_m \over \rho_c^2}
\biggl[1+\alpha_{dis} / (v\lambda)\biggr],
\label{rhoxx}
\end{equation}
and
\begin{equation}
\rho_{xy} = {\omega_c\rho_m \over \rho_c^2} = B/\rho_c.
\label{rhxy}
\end{equation}
Notice that $\rho_{xy}$ is not
modified to order $\alpha_{dis}$,
as was expected from the general analysis given earlier.
The above equations are the same as Eqs.~\ref{changerho1} to
\ref{changesig2}
if $\delta\lambda = \alpha_{dis}/v$.

The $v$ appearing in Eq.~\ref{v221} is the magnitude of the
average sliding velocity.
If one does not require self-consistency in the solution for the
sliding velocity in the presence of disorder potential,
one then must take $\vec v = \vec v_0$ from Eq.~\ref{field}.
However, Eq.~\ref{v221} can also be solved
self-consistently for the velocity, giving a self-consistent
$\vec v$.
These two procedures give the same result to order $\alpha_{dis}$,
but differ in the regime where the external field is not much
larger than the sliding threshold. The alternative
approach given by Eq.~\ref{addE} starts off with the self-consistent
velocity, and thus automatically gives results that are self-consistent, and
is in fact identical to what one obtains by solving Eq.~\ref{v221}
self-consistently.
However at fields not too far from the threshold field,
{\it a priori}, it is not clear which one gives the better solution.
We shall come back to this when we discuss
the Hall angle, and give more details for results at fields
not too larger than the threshold field where these
two treatments differ.
Our results for the conductivity and resistivity above
were given to order $\alpha_{dis}$, at which level the self-consistent
theory and the non-self-consistent theory agree.

Our results for the
conductivity and resistivity can be used to compare
with experiments.
In Fig.~1, we show the calculated $I-V$ characteristics of
a sliding Wigner crystal in the presence of disorder
scattering. We assume a Hall bar geometry of measurement, {\it i.e.},
we assume a fixed $I$, and plot the longitudinal ($V_{L}$) and
Hall ($V_{H}$) voltages as a function of $I$.
In Fig.~2, the differential resistivity is depicted
as a function of the applied voltage, also in the
Hall bar geometry. Our results are not expected
to be reliable close to depinning threshold. In particular,
the exponent of the $I-V$ curve given by the present perturbation
theory is incorrect near threshold. The dotted
portions of the lines are meant to convey the approximate nature of our results
in this regime. We shall compare these with the experimental
findings \cite{I,II,III,Jiang} later in this paper.

\subsection{Relation to the Static Depinning Field}

We here show that the quantity
$\rho_m\alpha_{dis} / \rho_c$ appeared in the last
subsection is
also a measure of the threshold field which
has been
defined in Eq.~2 in terms of
properties of the
static Wigner crystal. The static
$E_{Th}$, from Eq.~2, is \cite{NLM}:
\begin{eqnarray}
E_{Th} &\sim& {\rho_mc^Ta \over \rho_c\xi^2_T}, \nonumber \\
\xi_T &\sim& {{\rho_mc^Ta^2} \over {\sqrt{n_i}V}},
\label{ETH1}
\end{eqnarray}
where $n_i$ and $V = <V^2>^{1/2}$
are the average impurity density and strength.
Combining Eqs.~\ref{ETH1} and putting $c^T =\alpha\Omega^2a^2$,
we have
\begin{equation}
E_{Th} \sim {{n_iV^2/\rho_ca^3}   \over {\rho_m\alpha\Omega^2a^2}  }.
\label{ETH2}
\end{equation}

We may also estimate $E_{Th}$ as the field at which the second-order
correction to velocity becomes comparable to the zeroth-order
term. From Eq.~\ref{v221}, we estimate:
\begin{equation}
E_{Th} \sim {\alpha_{dis}\rho_m \over \rho_c}
=\sum_{\vec G}
{{|\rho (\vec G)|^2\Gamma (\vec G) / \rho_ca^3} \over
             {\rho_m\alpha\Omega^2a^2}}.
\label{ETH3}
\end{equation}
Indeed the above two expressions are
equivalent.
Therefore the high-velocity perturbation theory in fact
gives an estimate for the threshold field which can
be used to estimate other quantities such as the
correlation length as was done in Ref.~\cite{NLM}.
Using this, one may also relate the quantity $A(\vec G)$
defined earlier to the depinning threshold field. We then have:
\begin{equation}
{2 \over 3}\alpha_0 \sum_{\vec G} |\vec G|A(|\vec G|)
\stackrel{\sim}{=} \rho_c E_{Th}.
\label{AandETH}
\end{equation}

\subsection{Dependence of Hall Angle on Sliding Current}

As we mentioned already, the relationship between the external
field and the total sliding current depends on whether one requires
self-consistency in the velocity for fields not too large compared
to the depinning field.

In the case where we do require self-consistency in
Eq.~\ref{v221}, {\it i.e.\/}, the velocity that we
use on its right-hand-side
is not $\vec v_0$ in Eq.~\ref{field},
but the self-consistent $\vec v$, we find:
\begin{equation}
v = {
{-\alpha_{dis}\lambda +
\sqrt{\alpha_{dis}^2\lambda^2
+(\omega_c^2+\lambda^2)(\rho_c^2E^2/\rho_m^2-\alpha_{dis}^2)}
}
\over {\omega_c^2 + \lambda^2}
      }.
\label{vande}
\end{equation}
For $E \gg \alpha_{dis}\rho_m/\rho_c$, it reduces to
$\rho_cE/(\rho_m\sqrt{\omega_c^2+\lambda^2})$; and
for $E$ only slightly greater than
$\alpha_{dis}\rho_m/\rho_c$,
it is $(\rho_cE/\rho_m - \alpha_{dis})/\lambda$.
The Hall angle at an arbitrary velocity $v$ is:
\begin{equation}
tan\theta_H = {{\omega_c v} \over {\lambda v + \rho_m\alpha_{dis}/\rho_c}}.
\label{HAngle1}
\end{equation}
It becomes $45^o$, {\it i.e.,}
the field along the current becomes equal to the field
perpendicular to it, when:
\begin{equation}
\lambda + \alpha_{dis} / v = \omega_c.
\label{HallAngle45}
\end{equation}
Putting in the velocity in Eq.~\ref{vande}, substituting
$\alpha_{dis}$ with $\rho_cE_{Th}/\rho_m$
as we argued in the previous subsection,
and solving for $E$, we find:
\begin{equation}
{E_{45^o} \over E_{Th}} = \sqrt{2} {\omega_c \over \omega_c-\lambda}.
\label{HallAngle}
\end{equation}
For $E$ less than that given in Eq.~\ref{HallAngle},
the Hall angle is smaller than $45^o$ and the
field and the current are largely parallel. It becomes
greater than $45^o$ for greater $E$.
For $\omega_c$ less than $\lambda$, the Hall angle can never reach
$45^o$.

The results are different if one does not require self-consistency
in the velocity,
{\it i.e.\/}, if one uses $\vec v_0$
from Eq.~\ref{field} in
the right-hand-side of Eq.~\ref{v221}.
In this case, the velocity is related to the external field
as:
\begin{equation}
v = \sqrt{{  {E^2+E_{Th}^2-2EE_{Th}\lambda/\sqrt{\lambda^2+\omega_c^2}}
 \over {\lambda^2+\omega_c^2                                    }
    }},
\label{vande1}
\end{equation}
where we have used $E_{Th}$ for $\rho_m \alpha_{dis} / \rho_c$.
The large velocity limit remains the same, but in the limit
of $E$ only slightly greater than $E_{Th}$, it becomes,
\begin{equation}
v = E_{Th} \sqrt{   { 2(1-\lambda/\sqrt{\lambda^2+\omega_c^2})  \over
\lambda^2+\omega_c^2   }}.
\label{vande2}
\end{equation}
Now the Hall angle is:
\begin{equation}
tan \theta_H = { {\omega_cE-2\lambda\omega_cE_{Th}/\sqrt{\lambda^2+\omega_c^2}}
\over {\lambda E-(\lambda^2-\omega_c^2)E_{Th}/\sqrt{\lambda^2+\omega_c^2}} }.
\label{Hallangle2}
\end{equation}
It becomes $45^o$ when the total velocity is:
\begin{equation}
v_{45^o} = \sqrt{2} { {\omega_c E_{Th}} \over
(\omega_c-\lambda)\sqrt{\lambda^2+\omega_c^2} }.
\label{v45degree}
\end{equation}

We should add that these results must be treated with caution
at fields just above the threshold. Both
approaches given above are extrapolations of the perturbation
theory to a regime where perturbative
corrections are no longer small compared to the unperturbed term.
However, if we identify $E_{Th} = \alpha_{dis}\rho_m/\rho_c$,
then the self-consistent
solution is more sensible for applied fields not much greater
than
$E_{Th} = \alpha_{dis}\rho_m/\rho_c$, in that it gives a sliding velocity
$v \propto (E-E_{Th})$. Although the exponent relating the
velocity to the field is incorrect, $v$ does approach zero at the
depinning threshold field $E_{Th}$.

\section{Correlation Functions and Local Strain}

To better understand the state of a sliding Wigner
crystal, we consider in this section some correlation
functions. In charge-density waves, examination
of the strain has proven useful in understanding
the qualitative features of the Fukuyama-Lee-Rice model
\cite{SNC} and determining the limit of validity of the
elastic model.
We use the perturbation
theoretical analysis in Sec.~III to
study the correlation functions of the magnetically-induced Wigner crystal
under
the sliding condition.
We show our results for the velocity-velocity correlation
function in Sec.~IVA. A more detailed study of
the correlation length in the sliding
state is given in Sec.~IVB.
The results for the strain are given in Sec.~IVC.

\subsection{Correlation Functions}

In the sliding state, only the fluctuating parts of the displacement
and velocity need to be considered in the correlation functions.
We begin with the solution to Eq.~\ref{EOM1} for the displacement
in the DC limit:
\begin{equation}
\vec x(\vec k,t) = \int {{d^2\vec q} \over {(2\pi)^2}}
e^{i\vec q\cdot\vec v t}
(-\rho(\vec k-\vec q))\biggl[{\bf G}(\vec k,-\vec q\cdot \vec v)
\cdot (i\vec q)\biggr] V_{dis}(\vec q).
\label{xk}
\end{equation}
$\vec x(\vec k,t)$ is the Fourier transformation of $\vec x(\vec r,t)$.

After averaging over disorder, one has for $k \ll G$ for
equal-time correlation function:
\begin{equation}
< \vec x(\vec k,t) \vec x(-\vec k,t) > = \sum_{\vec G}
|\rho(\vec G)|^2\Gamma(\vec G)
\biggl[{\bf G}(\vec k, -\vec G \cdot \vec v) \cdot \vec G\biggr]
\biggl[{\bf G}(-\vec {k}, \vec G \cdot \vec v) \cdot \vec G\biggr],
\label{xxcor}
\end{equation}
and similarly for the velocity,
\begin{equation}
< {\vec v(\vec k,t)} {\vec v(-\vec k,t)} > = \sum_{\vec G}
|\vec G \cdot \vec v|^2|\rho(\vec G)|^2\Gamma(\vec G)
\biggl[{\bf G}(\vec k, -\vec G \cdot \vec v) \cdot \vec G\biggr]
\biggl[{\bf G}(-\vec {k}, \vec G \cdot \vec v) \cdot \vec G\biggr].
\label{vvcor}
\end{equation}

In perturbation theory
on the FLR model, Matsukawa and Takayama
\cite{MT} and Fisher \cite{DFisher}
have argued that the expression for the
positional correlation function contains divergent terms
at 4-$th$ order in the perturbation series, but the
expression for the velocity
correlation function does not up to the highest order terms examined.
For this reason, we view our velocity-velocity
correlation function as being more reliable. However, at the
present level of perturbation theory, the two correlation functions
behave in almost exactly the same way and are described by
the same correlation length that we discuss in the next subsection.

The correlations at large distance can be obtained by
the Fourier transformation of the above equations. The $\vec r$-dependence
of the real-space correlation functions is given by the
$\vec k$-dependent part of the $\vec k$-space correlation
functions:
\begin{equation}
g(\vec r) \sim \int {{d^2\vec k} \over (2\pi)^2} e^{i\vec k\cdot \vec r}
|{\bf G}(\vec k, -\vec G\cdot\vec v)|^2.
\label{gofr}
\end{equation}
The imaginary part of the poles in
the Green's function ${\bf G}$ determines the
characteristic lengths in these correlation
functions. For these dynamic quantities,
the magnetic
field necessarily enters in contrast to the static case.
We discuss these lengths in
the next subsection.

\subsection{Dynamic Correlation Length}

Before we describe our results for the correlation length which
is contained in Eq.~\ref{gofr},
it is useful to review the results for a CDW system treated
within the FLR model. The corresponding equation of motion was
given in Eq.~\ref{EOMFLR}.
In the CDW case, the Green's function
takes a simpler form:
\begin{equation}
{\bf G}_{CDW}(\vec k, \omega)
= {1 \over {\Delta k^2 - i \rho_m \lambda \omega}}.
\label{CDWG}
\end{equation}
Here $\Delta$ is the CDW elastic constant.

The correlation functions for the FLR model
in 3D have been calculated by Matsukawa and
Takayama
\cite{MT}. One can easily extend their result to 1D and 2D. We
find that the velocity-velocity correlation as defined in the
last subsection in the large
$r$-limit becomes
$({1 \over \sqrt{r}})^{D-1}e^{- {r \over \xi_v\sqrt{2}}}$
times a purely oscillatory factor
$sin({r \over{\xi_v\sqrt{2}}} - \phi_0)$,
where $\phi_0$ is a phase factor which depends on dimensionality $D$,
and the correlation length $\xi_v$ is given
by
\begin{equation}
\xi_v = \sqrt{\Delta /(\rho_m\lambda G|v_z|)}.
\label{CDWcor}
\end{equation}

This expression
for the correlation length holds as long as $\omega \ll \lambda$,
{\it i.e.,} as long as the $\omega^2$-term can be
ignored in the Green's function so that
the oscillation at $\omega$ is
purely diffusive.
$\xi_v$ is nothing but the decay length of an oscillation at
frequency $\omega = |\vec G \cdot \vec v|$.
It describes the distance over which a disturbance at
frequency $\omega_v$ will propagate.

This ``dynamic correlation length" can be related to the static
Fukuyama-Lee-Rice length $\xi$ in the CDW case as:
\begin{equation}
\xi_v = \xi \sqrt{\Delta \over {\xi^2 \rho_m \lambda  G|v_z|}}
\sim \xi \sqrt{E_{Th} \over {E-E_{Th}}}.
\label{lengthlength}
\end{equation}
So for fields $E$ not too large compared to the sliding threshold
field $E_{Th}$, the dynamic correlation length is comparable to
the static correlation length in the $B=0$ CDW case.

With this picture in mind, we examine the present situation which is
more complex. Since it is clear from the above
arguments that the dynamic correlation
length is determined by the excitations of a Wigner crystal
at the washboard frequency $\omega_v$, we examine the excitation spectra
which appear as poles in the Green's function,
given by Eq.~\ref{gpoles} and Eq.~\ref{kdefs}.
Setting the denominator of the Green's function to zero, one
has simply in terms of $\kappa_{\omega}$ and $\kappa_{\lambda}$:
\begin{equation}
(ka)^3 - \kappa_{\omega}^3 + i (ka) \kappa_{\lambda}^2 = 0.
\label{poles}
\end{equation}
We are interested in the case where $\kappa_{\omega}$ and
$\kappa_{\lambda}$ are given by those at
$\omega = \omega_v = |\vec G \cdot \vec v|$.

Note that Eq.~\ref{poles} is a cubic
equation for $k$. So for each given $\omega$, one finds generally
three poles
in the complex $k$-plane.
If $\omega_v$ is very small, then the modes are
once again diffusive. In this frequency range, there is no propagating
$q^{3/2}$-dispersive mode.
More quantitatively, at very small
average sliding velocity, for $\kappa_{\omega} \ll \kappa_{\lambda}$,
the three roots of Eq.~\ref{poles} are approximately:
\begin{equation}
k_{1,2}^2a^2 + i\kappa_{\lambda}^2 = 0,
\label{smallroots}
\end{equation}
and,
\begin{equation}
-\kappa_{\omega}^3 + ik_3a \kappa_{\lambda}^2= 0.
\label{smallroots1}
\end{equation}
These poles give rise to two lengths,
as:
\begin{equation}
\xi_{1,2} \sim {1 \over k_{1,2}} \sim {a \over \kappa_{\lambda}},
\label{smalllength}
\end{equation}
and,
\begin{equation}
\xi_3 \sim {1 \over k_3}
\sim a {\kappa_{\lambda}^2 \over \kappa_{\omega}^3},
\label{smalllength1}
\end{equation}

The largest one of the three lengths, $\xi_3$,
will determine the large-distance
behavior of the velocity-velocity correlation function, although
stronger correlations between the velocities of different
elements will set in when their distance becomes comparable
to $\xi_{1,2}$.
The first one of the two lengths ($\xi_{1,2}$)
in this over-damped limit
is similar to the CDW case, since:
\begin{equation}
{a \over \kappa_{\lambda}} = \sqrt{c^T/(\rho_m\lambda|\vec G\cdot\vec v|)}.
\label{aoverkl}
\end{equation}
In the limit of $E$ only slightly greater than $E_{Th}$, it
can be rewritten using the small $v$-limit of Eq.~\ref{vande} as:
\begin{equation}
{a \over \kappa_{\lambda}} = \xi^T \sqrt{E_{Th} \over {E-E_{Th}}}.
\label{aoverkl1}
\end{equation}
But in the regime where $E$ is much larger than $E_{Th}$
while $\kappa_{\omega}$
is still much less than $\kappa_{\lambda}$, we have instead:
\begin{equation}
{a \over \kappa_{\lambda}} = \xi^T \sqrt{\omega_cE_{Th} \over {\lambda E}},
\label{aoverkl2}
\end{equation}
which differs from the CDW result in Eq.~\ref{lengthlength}.

As the sliding velocity increases,
we get into the regime where $\kappa_{\omega} \gg \kappa_{\lambda}$,
The three roots are now approximately:
\begin{equation}
k_1a = \kappa_{\omega} + i {\kappa_{\lambda}^2 \over {3\kappa_{\omega}}},
\label{largeroots}
\end{equation}
and,
\begin{equation}
k_{2,3}a = \kappa_{\omega} e^{\pm i 2\pi/3}.
\label{largeroots1}
\end{equation}
In this case,
the $k_1$-mode is almost propagating.
The corresponding lengths are related to the static
correlation lengths as:
\begin{equation}
\xi_{1} \sim a{\kappa_{\omega} \over \kappa_{\lambda}^2},
\label{largelength}
\end{equation}
and,
\begin{equation}
\xi_{2,3} \sim {a \over \kappa_{\omega}}.
\label{largelength1}
\end{equation}

In this limit,
$\xi_1$ still describes the decay length of an oscillation at
frequency $\omega_v$, but the other two
describe the dispersive propagation
within a length that is smaller than the decay length. This is due to
the particular dispersion of the $k^{3/2}$-mode.
This phenomenon
does not appear in the usual CDW systems where excitations at $\omega_v$
are always overdamped as long as $\omega_v \ll \lambda$.
In the present case, the magnetic field makes it possible that
even for $\omega_v \ll \lambda$ in the sliding
state, there is still an almost
propagating mode. It also has
interesting effects on the strain in the sliding Wigner crystal, which
is the subject of the next subsection.

For a discussion of the crossover between the two regimes
in a typical experimental system,
we refer the reader to Sec.~VIIIB.
Under most experimental conditions where the DC current is
a few $nA$, $\kappa_{\omega}$ and $\kappa_{\lambda}$ are comparable.
Since their ratio has only a weak dependence on the sliding
velocity ($\sim~v^{1/6}$) for all cases
of practical interest, neither limit that we discussed
above for the velocity-velocity correlation length is
strictly applicable.

\subsection{Strain}

One can also calculate from
Eq.~\ref{xk} the local
longitudinal $(L)$ and
transverse $(T)$ strain in the sliding
Wigner crystal,
defined as:
\begin{equation}
E^s_L =
<(\nabla \cdot \vec x(\vec r, t))^2>,
\label{strainL}
\end{equation}
and,
\begin{equation}
E^s_T = <(\nabla \times \vec x(\vec r,t))^2>.
\label{strainT}
\end{equation}
Eq.~\ref{strainT} gives the
local elastic energy associated with the shear modulus;
and Eq.~\ref{strainL}
would be the local
elastic energy associated with the bulk modulus, if there
were no long-range Coulomb interaction.

{}From Eq.~\ref{xxcor}, we find that:
\begin{equation}
E^s_L =
\sum_{\vec G} |\rho(\vec G)|^2 \Gamma(\vec G)\int {d^2\vec k \over (2\pi)^2}
|\vec k \cdot {\bf G}(\vec k, -\vec G\cdot \vec v)\cdot \vec G|^2,
\label{Lstrain}
\end{equation}
and,
\begin{equation}
E^s_T =
\sum_{\vec G} |\rho(\vec G)|^2 \Gamma(\vec G)\int {d^2\vec k \over (2\pi)^2}
|\vec k \times ({\bf G}(\vec k, -\vec G\cdot \vec v)\cdot \vec G)|^2.
\label{Tstrain}
\end{equation}

$E^s_T$
can be transformed into
\begin{equation}
E^s_T = \sum_{\vec G}
|\rho(\vec G)|^2\Gamma(\vec G) |\vec G|^2
{1 \over 4\pi\rho_m^2} {1 \over {a^4\Omega^4\alpha^2}}
\int_0^{\pi} d(ka)
{
(ka)^5
\over
{((ka)^3-\kappa_{\omega}^3)^2+(ka)^2\kappa_{\lambda}^4}
}.
\label{ETS1}
\end{equation}
Once again, the result of this integral
depends on the relative size of $\kappa_{\omega}$ and $\kappa_{\lambda}$, or,
more physically, depends on if there is an almost propagating
mode at the washboard frequency.
If not, {\it i.e.\/}, for $\kappa_{\omega} \ll \kappa_{\lambda}$,
\begin{equation}
E^{strain}_{T} = \sum_{\vec G} |\rho(\vec G)|^2 \Gamma (\vec G) G^2
{1 \over {4\pi\rho_m^2(c^T)^2}}
ln {1 \over \kappa_{\lambda}}
= \beta {\alpha_{dis} \over \Omega^2\alpha a}
ln{1 \over \kappa_{\lambda}},
\label{Tstrain1}
\end{equation}
while for $\kappa_{\omega} \gg \kappa_{\lambda}$,
\begin{equation}
E^{strain}_{T} = \sum_{\vec G} |\rho(\vec G)|^2 \Gamma (\vec G) G^2
{1 \over {4\pi\rho_m^2(c^T)^2}}
(ln {1 \over \kappa_{\omega}} +
{\pi \over 3}{\kappa^2_{\omega} \over \kappa_{\lambda}^2})
= \beta {\alpha_{dis} \over \Omega^2\alpha a}
(ln {1\over \kappa_{\omega}}
+ {\pi \over 3}{\kappa^2_{\omega} \over \kappa_{\lambda}^2}).
\label{Tstrain2}
\end{equation}
Here $\beta$ is a dimensionless constant of order
unity, and $\alpha_{dis}$ is $\sim ~\rho_c E_{Th}/\rho_m$,
as was given in Eq.~\ref{v221}.
As we have mentioned already and will
discuss further in Sec.~VIIIB, the second term
on the right-hand-side in Eq.~\ref{Tstrain2} is smaller than
the first in the experimental systems.

The first term in Eqs.~\ref{Tstrain1} and \ref{Tstrain2}
diverges as $v \rightarrow 0$ as $ln{E-E_{Th} \over E_{Th}}$.
The second term
in Eq.~\ref{Tstrain2}, which is due to the $k_1$-mode
in Eq.~\ref{largeroots}, has no
counterpart in the usual charge-density
wave cases. This is due to the
fact that in the CDW cases, one only considers the strongly
diffusive mode as long as the washboard
frequency is smaller than
$\lambda$. In the present case, however, under the same condition,
there is the possibility of having a mode like
$k_1$ in Eq.~\ref{largeroots} whose real
part is much larger than its
imaginary part. It directly
reflects the fact that the $k_1$-mode is only weakly
damped, causing an almost resonant
absorption at $(\omega_v,k_1)$.

Within our high velocity
perturbation theory, the local transverse strain divergees weakly
as $v \rightarrow 0$, suggesting that the elastic medium
theory used here will break down before the pinning transition is
reached. However, it is possible
that higher order terms may change the form of the
divergence found in Eq.~\ref{Tstrain1} and Eq.~\ref{Tstrain2},
or eliminate it entirely.
In all cases $E^{strain}_L$
is smaller than $E^{strain}_T$ by at least a factor
of $(2\pi/\alpha)^2$, and does not contain
the divergences that we discussed above if the Coulomb force
is unscreened.

The possibility that the elastic model breaks down at small
velocity has been
discussed
by Coppersmith and Coppersmith and Millis
\cite{SNC} for the Fukuyama-Lee-Rice model without a magnetic field.
Using Eq.~\ref{Tstrain1} or Eq.~\ref{Tstrain2} and
typical material parameters for a FQHE device outlined later
in this paper, we find that the transverse strain at a typical
transport
current of $1nA$ is slightly less than 10\%; the longitudinal
strain is completely negligible.

The static correlation length is
determined by the shear modulus alone, because
the shear modulus is much smaller than the bulk modulus \cite{NLM}.
We see that in the sliding state, disorder scattering also
induces principally transverse distortions. This is confirmed in
both the study of the nonlinear conductivity in the previous
section, and here by examination of the strain distribution.
However for the velocity-velocity correlation length,
the magnetic field and the bulk modulus both enter.

\section{Free Carrier Effects: Large Velocity Limit}

In the previous sections,
we have studied some aspects of the
sliding motion of a Wigner crystal in the fractional
quantum Hall regime.
We now
consider the effects of
free carriers, which may be characterized by a separate
conductivity tensor, on the
sliding motion of the Wigner crystal. These free carriers may
be thermally excited, or due to sample inhomogeneities.
We treat them phenomenologically.

We assume that
the effects of the free carriers can be represented by
a
conductivity tensor whose diagonal and
off-diagonal elements are $(\sigma^F_{xx}, \pm\sigma^F_{xy})$.
Under the assumptions that

\noindent
1) there is no interconversion between the free carriers and the
Wigner crystal, and,

\noindent
2) the coupling between the free carriers and the
Wigner crystal is purely electrostatic,

\noindent
it can be
shown \cite{NLM} that longitudinal
component $c^L$ in the
Green's function becomes:
\begin{equation}
c^L k \rightarrow
\Omega^2 \biggl[
         {{2\pi ka} \over
          {1+ika{\sigma_{xx}^F\Omega \over \sigma_0\omega}}}
         +(1+\alpha)(ka)^2
         \biggr].
\label{NewCL}
\end{equation}
Here $\sigma_0 = a\epsilon\Omega/2\pi$, slightly
smaller than $e^2/h$ for a typical GaAs/AlGaAs modulation
doped FQHE sample.

It is therefore clear that the importance of free
carriers is measured by the magnitude of
the quantity $M_F = ka{\sigma_{xx}^F\Omega / \sigma_0\omega}$.
We are interested in the
range of $(\vec k,\omega)$ which corresponds to
that of the disorder scattering potential.
The characteristic frequency of the disorder potential
is $\omega_v \sim v/a$ in the sliding state of sliding velocity $v$.
For a current of a few $nA$, which is typical of
most experiments, $v$ is $\sim 10^{-2}~m/s$. This
yields $\omega_v \sim 10^6$ Hz, much smaller than
$\Omega$. This has two consequences. The first is that for such
frequencies ${\bf G}(\vec k,\omega)$ can be approximated by that
given by the elastic medium theory, which is true with or
without the free carriers. Secondly, combining this
with the phonon dispersion which determines the
characteristic value of $\vec k$, we conclude that $M_F \gg 1$
as long as
$\sigma_{xx}^F/\sigma_0$ of the free carriers
is greater than $10^{-3}$, as a conservative estimate \cite{footnote1}.

This range of $\sigma_{xx}^F$ in fact
covers most of the experimentally measurable range.
We therefore conclude
that for our purposes,
we only need to consider the case $M_F \gg 1$. We then have:
\begin{equation}
c^L k =
-i\omega \lambda_1
         +\Omega^2(ka)^2,
\label{NewCL1}
\end{equation}
with
\begin{equation}
\lambda_1 = 2\pi\Omega {\sigma_0 \over \sigma_{xx}^F},
\label{lambda1}
\end{equation}
and the corresponding Green's function is given by:
\begin{equation}
{\bf G}^{-1} (\vec k, \omega)/\rho_m = \Omega^2a^2k^2 {{\vec k\vec k}\over k^2}
            +\alpha\Omega^2a^2k^2\bigl(1-{{\vec k\vec k}\over k^2}\bigr)
                         - i\omega \lambda_1
                            {{\vec k\vec k}\over k^2}
                         - i\omega
\left(\matrix{\lambda & -\omega_c \cr \omega_c & \lambda \cr}\right).
\label{Gk11}
\end{equation}
Its determinant $P(\vec k,\omega)$ is
\begin{equation}
P(\vec k,\omega) = \alpha\Omega^4(ka)^4 - \omega^2
\biggl[\omega_c^2 + \lambda(\lambda_1+\lambda)\biggr]
-i\omega\Omega^2(ka)^2
\biggl[\alpha\lambda_1 + (1+\alpha)\lambda\biggr].
\label{Pkomega}
\end{equation}
{}From here most of the results from the previous
sections follow without major changes. We therefore only
list the results
in the rest of this section.

\subsection{Nonlinear Conductivity}

The perturbative calculation proceeds as in Sec.~III.
We find it convenient to define in the present case three
dimensionless quantities:
\begin{equation}
\kappa_1^2 = {|\omega_v|\omega_c \over \sqrt{\alpha}\Omega^2},
\label{k1}
\end{equation}
\begin{equation}
\kappa_2^2 = {|\omega_v| \over \Omega^2} (\lambda_1 + \lambda / \alpha),
\label{k2}
\end{equation}
\begin{equation}
\kappa_3^2 = {|\omega_v| \over \Omega^2}(\lambda_1 + 2\lambda),
\label{k3}
\end{equation}
where $\omega_v = \vec G \cdot \vec v$ is the sliding washboard
frequency, and $\kappa_3$ is less than $\kappa_2$. $\kappa_1 /a$
is the wavenumber at which the lower hybrid phonon mode,
which disperses like $\omega^2 = \alpha\Omega^4 (ka)^4/\omega_c^2$
in the present case, matches the washboard frequency, and
$\kappa_2$ and $\kappa_3$ measure the strength of dissipation
in the present context.

In the limit of $\omega_c \gg \lambda$ and $\lambda_1 \gg \lambda/\alpha$,
we find that the spatially averaged velocity correction to second
order in disorder scattering $\vec v_2$ is:
\begin{equation}
\left(\matrix{\lambda & -\omega_c \cr \omega_c & \lambda \cr}\right)
\cdot \vec v_2 =
- {1 \over 16\alpha\Omega^2a^2\rho_m^2}
\sum_{\vec G} |\rho(\vec G)|^2\Gamma(\vec G)|\vec G|^2
\vec G ~sgn(\vec G\cdot\vec v) = -\alpha_{dis} {\vec v \over v}.
\label{fv2}
\end{equation}

This yields a resistivity tensor which is identical to that
in the absence of free carriers.
This is because the leading corrections depend solely
on transverse distortions,
and only the longitudinal modes are
affected by free carrier screening.
The results were given in Sec.~III, and will not be repeated here.
At higher orders of the perturbation theory, the longitudinal
channel affects the results; and the conductivities with or
without the screening effects will be different.

\subsection{Correlation Functions}

As in Sec.~IVB,
the correlation length
of the sliding state for all of the correlation functions
will be determined by the poles of the Green's function
${\bf G}(\vec k,\omega)$, given by:
\begin{equation}
(ka)^4 - \kappa_1^4 + i(ka)^2\kappa_2^2 = 0.
\label{correlength}
\end{equation}
For simplicity, we assume $\kappa_1 \gg \kappa_2$, and obtain:
\begin{equation}
\xi  = 4a \kappa_1 / \kappa_2^2 \sim 1/\sqrt{|\omega_v|},
\label{length}
\end{equation}
which approaches infinity
as $|\omega_v| \rightarrow 0$.

The evaluation for the strain also proceeds as before. We here only
give the results:
\begin{equation}
E^S_T = |\nabla \times \vec x|^2
= \sum_{\vec G} |\rho (\vec G)|^2\Gamma (\vec G)
{{|\vec G|^2} \over {4\pi\alpha^2\Omega^4a^4\rho_m^2}}
\int (ka)^3d(ka)
{ {(ka)^4 + \alpha \kappa_1^4(1+\lambda_1^2/\omega_c^2)}
\over {((ka)^4-\kappa_1^4)^2 +(ka)^4\kappa_2^4} },
\label{ETS}
\end{equation}
and
\begin{equation}
E^S_L = (\nabla \cdot \vec x)^2
= \sum_{\vec G} |\rho(\vec G)|^2\Gamma(\vec G)
{|\vec G|^2 \over 4\pi\alpha^2\Omega^4a^4\rho_m^2}
\int (ka)^3d(ka){{\alpha^2(ka)^4 + \alpha \kappa_1^4}
 \over {((ka)^4-\kappa_1^4)^2 +(ka)^4\kappa_2^4} }.
\label{ELS}
\end{equation}

We see once again that $E^S_T \gg E^S_L$, although the
actual ratio depends on the details of the relevant parameters.
For $\kappa_1 \gg \kappa_2$, we have:
\begin{equation}
E^S_T
= \sum_{\vec G} |\rho(\vec G)|^2\Gamma(\vec G)
{|\vec G|^2 \over 4\pi\alpha^2\Omega^4a^4\rho_m^2}
ln{1 \over \kappa_1} = {\beta\alpha_{dis} \over \Omega^2\alpha a}
                  ln{1 \over \kappa_1}.
\label{fETS1}
\end{equation}
Here $\beta$ is once again a constant of order unity.
The logarithmic divergence has the same origin as that for the
transverse strain distribution function previously. It is the result
of the dimensionality of the present problem.

\section{AC + DC Interference Effects}

We now consider the possible interference
effects due to the presence of both a DC field and an AC field.
It is more convenient here to adopt the
alternative
approach mentioned in Sec.~IIIB and determine the
electric field $\vec E^{Tot}$ required to
sustain a current density $\vec j$ which has
both DC and AC components, given by:
\begin{equation}
\vec j = \rho_c \vec v_{DC} + \rho_c \vec v_{AC} cos \omega_{AC} t,
\label{ACDCCurrent}
\end{equation}
within the perturbation theory,
\begin{equation}
\vec {E^{tot}} = \vec E + \vec {E^p},
\label{ACDCField}
\end{equation}
where
\begin{equation}
\rho_c \vec E = \rho_m
\left(\matrix{\lambda & -\omega_c \cr \omega_c & \lambda \cr}\right)
(\vec v_{DC} + \vec v_{AC} cos \omega_{AC} t),
\label{ACDCfield1}
\end{equation}
and to second order in disorder potential
\begin{eqnarray}
\rho_c \vec {E^p} & = & - i \sum_{\vec G, n_1, n_2} |\rho(\vec G)|^2
\int {d^2\vec q \over (2\pi)^2} \vec q \Gamma(\vec q)
J_{n_1}(\vec q \cdot \vec v_{AC} / \omega_{AC})
J_{n_2}(\vec q \cdot \vec v_{AC} / \omega_{AC}) \nonumber \\
                 &   &
e^{i\omega_{AC}(n_2-n_1)t}
\biggl(\vec q \cdot {\bf G}(-\vec q + \vec G, n_1\omega_{AC}
+ \vec q \cdot \vec v_{DC})
              \cdot \vec q
\biggr).
\label{acdc}
\end{eqnarray}
The notations here are the same as before.
$J_n(x)$ is the $n$-th
order integer Bessel function. The sums over $n_1, n_2$ include
both positive and negative integers.
The derivation of Eq.~\ref{acdc} is the same as its counterpart without
the AC field and essentially follows
that in {\it e.g.}
Ref.~\cite{Sneddon}. Therefore it
will not be repeated here.

There are two inteference effects:

1). For $n_1 = n_2$, the resultant DC field is modified
by the AC current (Shapiro anomaly);

2). The linear AC-response,
{\it i.e.\/},
the $\omega_{AC}$-component of $\vec {E^p}$, will be affected by the
DC current.

Both effects enter through the frequency dependence of the
Green's function ${\bf G}$ where $n\omega_{AC} + \vec q\cdot \vec v_{DC}$
appears. We discuss them separately in the next two subsections.

\subsection{Shapiro Anomaly }

Setting $n_1 = n_2$ in Eq.~\ref{acdc}, we have:
\begin{equation}
\rho_c \vec {E^p} = - i \sum_{\vec G, n} |\rho(\vec G)|^2
\int {d^2\vec q \over (2\pi)^2} \vec q \Gamma(\vec q)
\bigl(J_{n}(\vec q \cdot \vec v_{AC} / \omega_{AC})\bigr)^2
\biggl(\vec q \cdot {\bf G}(-\vec q + \vec G, n\omega_{AC}
+ \vec q \cdot \vec v_{DC})
              \cdot \vec q
\biggr).
\label{shapiro}
\end{equation}
The $n = 0$ term in the above equation gives us the
nonlinear conductivity in the absence of the AC field, which was
studied in previous sections. If $n \ne 0$, under the condition
that
$|\vec G \cdot \vec v_{DC} + n\omega_{AC}| / |\vec G \cdot \vec v_{DC}|
\gg max(\kappa_{\lambda},\kappa_{\omega})$, {\it i.e.,}
we are somewhat away from the resonance,  we find:
\begin{equation}
\rho_c \vec {E^p} =
\alpha_0
\sum_{\vec G, n} \vec G
sgn(\vec G \cdot \vec v_{DC} -n \omega_{AC} )
J^2_n(\vec G \cdot \vec v_{AC} / \omega_{AC})
A(\vec G).
\label{shapiro1}
\end{equation}
$A(\vec G)$ here is the same as
that given in Eq.~\ref{defAG}.

We here fix $\omega_{AC}$ to be positive, and have used the
fact that in summing over $\vec G$, $\pm \vec G$ enter together.
$\alpha_0$ in Eq.~\ref{shapiro1} is once
again $\pi/3$ if $\kappa_{\omega} \gg \kappa_{\lambda}$ and
$\pi/4$ if $\kappa_{\omega} \ll \kappa_{\lambda}$.

If we further relate $A(\vec G)$ to the threshold field
$E_{Th}$ through $\alpha_{dis}$ as was done earlier in
the paper in Eq.~\ref{defAG} and Eq.~\ref{AandETH},
we finally have
for a DC current $\vec j = \rho_c \vec v_{DC}$, the total external
DC field is:
\begin{eqnarray}
\rho_c \vec E = \rho_m
\left(\matrix{\lambda & -\omega_c \cr \omega_c & \lambda \cr}\right)
\cdot \vec v & + & \rho_c E_{Th} \vec v_{DC} / v_{DC} \nonumber \\
             & + &
 {\alpha_0\over 4}\sum_{\vec G,n} |\vec G| A(\vec G)
J_n(\vec G \cdot \vec v_{AC}/\omega_{AC})
sgn(|\vec G \cdot\vec v_{DC}| - n\omega_{AC}) \vec v_{DC} / v_{DC}.
\label{shapiro2}
\end{eqnarray}
In cases when
$\vec G$ is not parallel to $\vec v_{DC}$, as long as
$\vec v_{DC}$ is along one of the six nearest neighbor or the
six next nearest neighbor $\vec G$-vectors, summing over
all $\vec G$'s with the same $\vec G \cdot \vec v_{DC}$ would
yield a correction to the field in the direction of $\vec v_{DC}$.

We therefore see that the Hall resistivity is once again not altered.
The longitudinal resistivity shows upward jumps with increasing
current, or downward jumps with increasing frequency of the
AC field.
If one is instead interested in the conductivity, the direction
of the jumps will depend on whether or not the Hall angle is greater than
$45^o$. If not, downward jumps in resistivity will
translate into upward jumps in conductivity; otherwise, downward
jumps in resistivity will become a downward jump in conductivity.
This is familiar in experiments in the FQHE
regime \cite{FQHE0} where one often encounters a situation in which
the off-diagonal elements of the conductivity/resistivity tensor
are greater than the diagonal elements.
The maximum value of the $\vec G$-vectors that
can give rise to the jump is cutoff by either the $\vec G$-dependence
of disorder correlation function $\Gamma(\vec G)$ or by
the density $\rho(\vec G)
\sim e^{-|\vec G|^2 l_B^2/2}$. Both enter through the expression
for $A(\vec G)$.

This is the Shapiro step in the DC voltage, on
the order of the depinning field but smaller
in magnitude, appearing when we vary
either the frequency of the AC field or the
DC current. ($\rho_c \vec {E^p}$ is proportional to
the DC voltage.) The main experimental
signatures are therefore:

1). If one sweeps the frequency while keeping the
DC current constant, as $n\omega_{AC}$ crosses
$|\vec G \cdot \vec v_{DC}|$ from below,
the DC voltage decreases, leading to a decrease in
the DC resistivity.
The jump in
DC voltage should be proportional to the threshold field,
with the proportionality coefficient being
$J^2_n(\vec G \cdot \vec v_{AC}/\omega_{AC})$, which gives
its dependence on the amplitude and the frequency (or the DC current) of
the AC field.

2). If one instead gradually increases
the DC current at a fixed AC frequency, one observes an increase in
resistivity.

3). In usual CDW systems, one often measures
the differential resistivity and looks for
peaks at resonance. The width of the peak gives additional
information on the correlation length. In the presence case,
the situation is more complicated and the width will
depend on
the velocity-velocity correlation length which we
discussed extensively earlier.

Overall, neither $\rho (\vec G)$ nor $\Gamma(\vec G)$
decays
very fast with $\vec G$ for small filling factors with
small magnetic length. There will
be many more peaks than in the traditional CDW systems where
$\rho(\vec G)$ is finite at a single $\vec G$ and its inversion.
The features, if indeed observable, are expected to be
much broader because the contributions come
from many $\vec G$'s. Perhaps
a more severe hindrance is that the correlation length
in the Wigner crystals around $\nu = 1/5$ appears
short \cite{NLM}.
Traditionally, it has been easier to observe the
interference effect in the AC response. We turn to it
in the next subsection.

\subsection{Linear AC Response Function }

To obtain the linear AC response in the presence of a DC current,
we assume small $\vec v_{AC}$ and only keep terms linear in
$\vec v_{AC}$ in Eq.~\ref{acdc}. We have:
\begin{eqnarray}
\rho_c \vec {E^p} & = &-{i \over (2\pi)^2}e^{-i\omega_{AC} t}
\sum_{\vec G} |\rho(\vec G)|^2 \nonumber \\
                  &   &
\int d^2{\vec q} \vec q\Gamma(\vec q) (\vec q\cdot\vec v_{AC} / 2\omega_{AC})
\biggl( \vec q \cdot \bigl(
-{\bf G}(-\vec q + \vec G, \vec q \cdot \vec v_{DC}) +
 {\bf G}(-\vec q + \vec G,
\omega_{AC} + \vec q \cdot \vec v_{DC})
        \cdot \vec q \bigr)
\biggr) \nonumber \\
                 &   &
+ c.c.,
\label{acresponse}
\end{eqnarray}
from which we can determine the in-phase and the out-of-phase
component of the AC resistivity.
For the in-phase part that is proportional to $cos\omega_{AC} t$, we find:
\begin{equation}
(\rho_c\vec {E^p})_{in} = \alpha_0
\sum_{\vec G} \vec G sgn(\omega_{AC} - \vec G \cdot \vec v_{DC})
(\vec G \cdot \vec v_{AC} / \omega_{AC})
A(\vec G).
\label{acresponse1}
\end{equation}
And for the out of phase part that is proportional to
$sin\omega_{AC} t$, we have:
\begin{equation}
(\rho_c\vec {E^p})_{out} =
\sum_{\vec G} \vec G A(\vec G)
(\vec G \cdot \vec v_{AC} / \omega_{AC})
ln{ {max(\kappa_{\omega'},\kappa_{\lambda'})} \over
    {max(\kappa_{\omega},\kappa_{\lambda}}) }.
\label{acresponse2}
\end{equation}
where $\kappa_{\omega}$ and $\kappa_{\lambda}$ are defined as before and
the primed quantities are obtained by replacing $\vec G \cdot \vec v_{DC}$
with $\omega_{AC} - \vec G \cdot \vec v_{DC}$ in the respective unprimed
quantities.
We have
in terms of the threshold field:
\begin{equation}
(\rho_c\vec {E^p})_{in} = \rho_cE_{Th}
\sum_{\vec G} \vec G /|\vec G| sgn(\omega_{AC} - \vec G \cdot \vec v_{DC})
(\vec G \cdot \vec v_{AC} / \omega_{AC}),
\label{acresponse3}
\end{equation}
and
\begin{equation}
(\rho_c\vec {E^p})_{out} = \rho_c E_{Th} /4\alpha_0
\sum_{\vec G} \vec G / |\vec G|
(\vec G \cdot \vec v_{AC} / \omega_{AC})
ln{ {max(\kappa_{\omega'},\kappa_{\lambda'})} \over
    {max(\kappa_{\omega},\kappa_{\lambda}}) },
\label{acresponse4}
\end{equation}
where the sums are over those that give the
same $\vec G \cdot \vec v$ among the six smallest $\vec G$'s
only.
The above equations also hold for larger $\vec G$'s, with the
necessary replacement of $E_{Th}$ in Eq.~\ref{acresponse3}
by an expression in which the
$|\vec G|$-dependence of various terms in $A(\vec G)$ is restored.

The out-of-phase part of the AC resistivity should show an inductive to
capacitive transition as $\omega_{AC}$ goes from below
$\sim \vec G \cdot \vec v_{DC}$ to above. For large
$\omega_{AC}$, it becomes ${1 \over \omega_{AC}}ln \omega_{AC} $.
We also note that it
has only a component along the direction of $\vec v_{DC}$, that is,
it only enters the longitudinal resistivity.

The in-phase term involves now three vectors: $\vec G$,
the DC current $\vec v_{DC}$, and the AC current $\vec v_{AC}$. In cases
when the latter two are not aligned, it will enter both the
diagonal and off-diagonal AC resistivity. In particular,
if the relevant $\vec G$ satisfies $\vec G \cdot \vec v_{AC} = 0$,
the jump in the real part of the
AC resistivity disappears, although the imaginary
part will still show an inductive anomaly.
If $\vec v_{AC}$ is parallel to $\vec v_{DC}$, then
\begin{equation}
(\rho_c\vec {E^p})_{in} = \rho_cE_{Th}
sgn(\omega_{AC} - |\vec G \cdot \vec v_{DC}|)
/(a \omega_{AC}) \vec v_{AC}.
\label{acresponse5}
\end{equation}
The equation is uncertain up to a numerical factor
smaller than unity.
The diagonal resistivity changes from roughly
${\rho_m \over \rho_c^2} (\lambda - \alpha_{dis} /a\omega_{AC})$
to
${\rho_m \over \rho_c^2} (\lambda + \alpha_{dis} /a\omega_{AC})$
as $\omega_{AC}$ increases through $\vec G \cdot \vec v_{DC}$.
Approximately, for the out-of-phase part, we find:
\begin{equation}
(\rho_c\vec E)_{out}
= \sum_{\vec G} A(\vec G) {{\vec v_{AC}} \over a\omega_{AC}}
ln{{|\omega_{AC} - \vec G \cdot \vec v_{DC}|} \over {\vec G\cdot \vec v_{DC}}}.
\label{acresponse6}
\end{equation}

We also note that when the DC current $\vec v_{DC}$ and
the AC current $\vec v_{AC}$ are not parallel, the system no longer
exhibits the rotational invariance that we invoked in arguing for
a classical Hall resistivity. Indeed, in this case, we find that
the Hall component of the in-phase AC resistivity will not
be $B/\rho_c$, with corrections occurring first in second order in
disorder potential.

Care must be exercised when converting these statements about the
changes in the AC resistivity
to those in conductivity, as in
the case of the Shapino anomaly that
we discussed in the last subsection. When the Hall angle for the
AC current is smaller than $45^o$, upward jumps in
the real part of the
diagonal resistivity translates into downward jumps in
the real part of the diagonal conductivity, and the inductive to capacitive
transition remains. But for Hall angles greater than
$45^o$, the situation will be reversed.
we therefore choose to give us results in terms of
resistivity.

\section{Results for Zero Magnetic Field}

In the case of a zero magnetic field, the Green's function has
the same form, with $\omega_c$ now equal to 0,
and $P(\vec k,\omega)$
replaced by:
\begin{equation}
P(\vec k,\omega) = 2\pi\alpha\Omega^4\rho_m
\bigl[ (ka)^3 - \kappa_{\omega}^3 - i\kappa_{\lambda}^2 (ka) sgn(\omega)\bigr],
\label{zerofield1}
\end{equation}
where as before, we have ignored the $\alpha (ka)^2 / a\pi$ term
in the last term on the right-hand-side, and
\begin{equation}
\kappa^2_{\lambda} = {{|\omega| \lambda} \over {\alpha\Omega^2}},
\label{zerofield2}
\end{equation}
and,
\begin{equation}
\kappa^3_{\omega} = \kappa^2_{\lambda} {{|\omega| \lambda} \over
{2\pi\Omega^2}}
 = {{\omega^2 \lambda^2} \over {2\pi\alpha\Omega^4}}.
\label{zerofield3}
\end{equation}

All of the results for the resistivity are in exactly the same
form as those
of the diagonal resistivity with the magnetic field,
if
we set $\omega_c = 0$ and put in the new $\kappa_{\lambda}$
and $\kappa_{\omega}$.
We will not repeat them here.
The only exceptions are the velocity-velocity correlation
length. We give the new results here:

1). For $\kappa_{\omega} \ll \kappa_{\lambda}$, {\it i.e.}, the
overdamped case,
\begin{equation}
\xi_{1,2} = \xi^T \sqrt {E_{th} \over E-E_{Th}},~~~~~~
\xi_3 = \xi^L {E_{th} \over E-E_{Th}};
\label{zerofield4}
\end{equation}

2). For $\kappa_{\omega} \gg \kappa_{\lambda}$,
\begin{equation}
\xi_{2,3} = \xi^T\bigl({\xi^L \over \xi^T}\bigr)^{1/3}
({E_{Th} \over E-E_{Th}})^{2/3},~~~~~~
\xi_1 = \xi_T\bigl( {\xi^T \over \xi^L} \bigr)^{1/3}
\bigl({E_{th} \over E-E_{Th}}\bigr)^{1/3}.
\label{zerofield5}
\end{equation}

In the overdamped region as specified in case 1, the
equations are the same as for the CDWs. Case 2 is
where the differences lie.

\section{Summary of Our Main Results}

In Sec.~VIIIA, we summarize the main theoretical
results regarding the
transport properties of a sliding Wigner crystal
obtained in this work.
In Sec.~VIIIB, we estimate various characteristic lengths and frequencies
that have appeared in our theory using parameters appropriate for a
typical high mobility modulation doped GaAs/AlGaAs
heterojunction.
Our calculations should be strictly valid only in the very low
temperature limit, while experiments are done at
a variety of temperatures. Therefore
in Sec.~VIIIA, we discuss the
possible temperature effects
on the nonlinear $I-V$ curves.

\subsection{Summary of the Theoretical Transport Properties}

We summarize our main results in the form of predicted $I-V$
characteristics of a sliding Wigner crystal.

\vskip 1.0truecm
\noindent
A). Hall bar geometry

In the Hall bar geometry, one controls the current that runs
through the sample.
If $\vec j = \rho_c \vec v$
along the $x$-direction,
then,
\begin{equation}
E_x
= {\lambda\rho_m \over \rho_c^2}
j_x + E_{Th},
\label{eandj1}
\end{equation}
\begin{equation}
E_y = {\omega_c\rho_m \over \rho_c^2} j_x.
\label{eandj2}
\end{equation}

We notice that
the functional form of Eq.~\ref{eandj1}
is in very good agreement with the experiments by Li
{\it et al.} \cite{III} and those by Williams {\it et al.}
\cite{II}. It also agrees with the large
depinning field data of Jiang {\it et al.} ($V \sim 50 mV$) \cite{Jiang}.
(See Fig.~8 in their paper.)
We discuss the comparison in more detail in Sec.~IX.

The first equation holds also for the case of a sliding Wigner crystal
without a magnetic field, in which case there is simply no Hall component.
The same equations also hold when there is coupling to
free carriers as long as we are in the large velocity regime.
Figs.~1 and 2 given previously in this paper depict the theoretical
results in the Hall bar geometry.

\vskip 1.0truecm
\noindent
B). Corbino geometry

In Corbino geometry, the electric field is controlled by virtue of
symmetry of the experimental set-up.
If a constant DC field $\vec E$ is along
the $x$-direction, {\it i.e.\/},
$E_y =0$,
then for $\omega_c \gg \lambda$,
\begin{equation}
j_x = {\rho_c^2 \over \rho_m}{\lambda \over \omega_c^2}
(E_x + {\alpha_{dis}}{\rho_m \over \rho_c}{\omega_c \over \lambda}),
\label{jxande}
\end{equation}
and
\begin{equation}
j_y = {\rho_c^2 \over \rho_m}{1\over\omega_c}
(E_x - 2{\alpha_{dis}}{\rho_m \over \rho_c}{\lambda\over\omega_c}).
\label{jyande}
\end{equation}
And the total current $j$ is given to order $\alpha_{dis}$
by:
\begin{equation}
j = {\rho_c^2 \over \rho_m}{1\over\omega_c}
(E-E_{Th}{{\lambda} \over {\omega_c}}).
\label{jtotale}
\end{equation}

\vskip 1.0truecm
\noindent
C). Hall Angle

The Hall angle, {\it i.e.,} the angle
between the current and the external field, increases
with the external field. In the theory
in which the velocity on the right-hand-side
of Eq.~\ref{v221} is the self-consistent velocity, taking into account
the disorder effects on the velocity,
rather than the bare $\vec v_0$ in
Eq.~\ref{field}, the Hall angle becomes
$45^o$ when $E/E_{Th} \stackrel{\sim}{=}
\sqrt{2} \omega_c / (\omega_c - \lambda)$.

\vskip 1.0truecm
\noindent
D). Shapiro anomaly

The longitudinal DC field changes by
$\sim E_{Th} J_n(\vec G\cdot
\vec v_{AC}/\omega_{AC})
sgn(|\vec G\cdot \vec v_{DC}| - n\omega_{AC})$ across
the resonance as the DC current or the AC frequency
is swept.

\vskip 1.0truecm
\noindent
E). AC response

The out-of-phase AC response only occurs in
the longitudinal resistivity. It changes from
inductive to capacitive as the frequency increases
through resonance.

The in-phase AC field will change by
$\sim E_{th}
sgn(\omega_{AC} - |\vec G \cdot \vec v_{DC}|)
|\vec G \cdot \vec v_{AC}|/\omega_{AC})$
across the resonance. If the DC current and the AC
current are not in the same direction, it will have
both diagonal and off-diagonal components.

\subsection{Characteristic Lengths and Frequencies}

The elastic medium theory that we adopted to describe
the sliding Wigner crystal is only valid in the long
wavelength limit. In this subsection, we estimate various
characteristic length scales and frequency scales that enter
our previous analysis. We use numbers that are typical of
a high mobility modulation-doped GaAs/AlGaAs heterojunction.
Therefore $m = 0.067~m_e$, $\epsilon = 13$, electron density
$n = 10^{11}~cm^{-2}$, magnetic field $B = 20~Tesla$,
mobility $\mu = 10^{7}~cm^{2}/V \cdot s$,
and when needed, the size of the sample
is assumed to be $3~mm~\times~3~mm$. The ratio of the shear
modulus versus the bulk modulus $\alpha$ is taken to be
its classical zero-temperature value of 0.02 \cite{BM77}. Of these
parameters,
$\alpha$ has the largest uncertainty, with
corrections from
both quantum and thermal effects, the former being
perhaps more important near the zero-temperature
Wigner solid/FQHE liquid boundary, the latter
more important near the thermal melting transition
of the Wigner solid.

With these numbers, the cyclotron frequency
$\omega_c \sim 10^{13}s^{-1}$,
$\Omega \sim 10^{12}s^{-1}$. At $qa = 1/3$,
the zero-field phonon frequencies are: $\omega_L \sim 10^{12}s^{-1}$,
$\omega_T \sim 10^{11}s^{-1}$, and the acoustic
magnetophonon frequency is
$\sim 10^{10}s^{-1}$.
The collision frequency deduced from the zero-field
mobility is $\tau^{-1} \sim 10^{9} s^{-1}$. How this number
should change in a strong magnetic field is
an open question, as is the relation between $\tau^{-1}$ and the
parameter $\lambda$ in Eq.~\ref{EOM}. Values of
$\lambda$ inferred from the high
field $I-V$ measurements are much larger,
$\lambda \sim 10^{12} - 10^{13} s^{-1}$.

Typically, measurements are done with a current $I \sim 1nA$
in the sliding regime. Assuming a large fraction of the
Wigner crystal is sliding, we have $v \sim 10^{-2} ms^{-1}$.
This gives a washboard frequency of
$\omega_v$ between $10^5$ to $10^6~s^{-1}$.
We see that this is much smaller than the characteristic phonon
frequencies involved. Therefore, the elastic medium theory should
be valid. We also have $\kappa_{\omega} \sim 10^{-4}$, implying a typical
wavelength $10^{4}a$, which approaches the typical size of a sample.
It has been shown that within the perturbation theory for a CDW
system \cite{Sneddon}, the broad band noise vanishes upon average
over volume in an infinite system, but does not vanish if the system size is
not infinite.
The fact that $\kappa_{\omega}$ is comparable to
system size in the present case may be responsible for
the sizable broad
band noise observed experimentally \cite{I,III}. It would be very interesting
to see if experimentally the narrow band noise around $\omega_v$
could be observed.
$\kappa_{\lambda}$ is between $10^{-3}$ to $10^{-4}$,
depending on the size of $\lambda$ chosen.

We note that if one converts the resistivity in the sliding
state as measured by the experiments \cite{I,II,III,Jiang}
to a scattering frequency using Eq.~\ref{addEsol} in the high
field limit,
the scattering frequency $\lambda$ is rather large.
In fact, it is comparable to $\omega_c$, the largest frequency scale
in the problem. It is much larger than the scattering rate deduced
from the zero-field mobility, and much larger than
that deduced from the metallic state resistivity
at larger filling factors than $\nu = 1/5$ \cite{Jiang1}.
At present, it is unclear what mechanism is responsible for such
a high resistivity in the sliding regime.

We suggest that it
is, at least at
moderately high temperatures, due to the damping caused by the
backflow of free carriers to screen out the charge
density fluctuations in a sliding Wigner crystal \cite{PBL}.
It is clear from Eqs.~\ref{NewCL} and \ref{NewCL1}
and the ensuing discussion that
the free carriers lead to a longitudinal mode
dissipation proportional to the free carrier resistivity.
Although in the present problem, this does not enter the
nonlinear conductivity to second order in
the randomness, it presumably will in higher orders, as in the CDW problem
\cite{PBL}.

\subsection{Nonzero Temperatures}

The calculations reported in this paper did not include thermal
effects explicitly, while at least in some experiments \cite{Jiang}
such effects are quite apparent. In this subsection, we outline the
effects that we expect to occur at $T > 0$.

We believe the four important effects occurring as $T$ increases
are: 1) an increase in the number of free carriers; 2) a decrease
in the strength of the interaction of the Wigner crystal with the impurity
potential; 3) an increase in the number as well as in
the size of dislocation pairs; and
4) the subsequent melting of the Wigner crystal, presumably via a
Kosterlitz-Thouless transition \cite{KT}.

The first point is straightforward: presumably, creation of free carriers
involves excitations over an energy gap, $\Delta_g$, and so
the number of free carriers and thus the free carrier conductivity should
have a roughly activated temperature dependence. One therefore
expects the conductivity at $E < E_{Th}$ to be activated.
Interestingly, in at least some experiments \cite{Jiang} the conductivity
at $E > E_{Th}$ is also activated, with the same gap, suggesting that the
conductivity at $E > E_{Th}$ is also controlled by the free carriers.
An additional point, inferred from the experiments in \cite{Jiang},
is that there is one single excitation gap at temperatures below
$\sim 140 mK$. In other words, the low-energy charged excitations in
the Wigner crystal appear to be well-defined, and well-separated from
higher energy excitations.

Free carriers not only provide conductivity and dissipation, they may
also screen the impurity potential. Also
thermal fluctuations of the Wigner crystal
may tend to average out the impurity potential. For these reasons, we
expect that the effective strength of the
random potential will get weaker as the temperature is raised.

The third effect of finite temperatures, the presence and
the proliferation of dislocation pairs, requires mode discussion. The
conventional
Fukuyama-Lee-Rice \cite{FLRMod} theory of pinning and nonlinear conductivity
is based on a ``phase-only" model in which the
amplitude flutuations of the order-parameter or the dislocations are
not allowed. One important consequence of forbidding
the formation of
dislocations is the existence of a well-defined threshold field
below which the whole solid is pinned and above which the whole solid
moves collectively. In a model in which
dislocations are allowed, it is possible for some portions of the solid to move
while other portions remain pinned at any given moment; and
the depinning transition is no longer
sharp nor is it unambiguously defined \cite{SNC}.
In a two-dimensional ordered solid, the elastic
energy of an isolated dislocation
is logarithmically divergent in system size so at low temperatures
the dislocations are created via thermal activation of bound pairs
whereby the divergent elastic energy due to each one cancels out.
As $T$ increases, the number of pairs increases as does the typical size of
a pair, $\xi_{KT}$. Dislocation pairs proliferate and $\xi_{KT}$
diverges as the Kosterlitz-Thouless melting temperature is
approached from below \cite{KT}.

Another way to see this is as follows. At zero or very low
temperatures, disorder pinning dominates over the thermal effects and
the Fukuyama-Lee-Rice theory \cite{FLRMod} gives the disorder-induced
correlation length $\xi_T$ beyond which the positional order
disappears. As temperature is raised, $\xi_T$ decreases according to
$\xi_T \propto c^T$, proportional to
the (decreasing) shear modulus, assuming that the change in disorder
potential itself is smooth and featureless. However the increasing
thermal fluctuations also give rise to a thermally-induced correlation length
$\xi_{KT}$ on the order of the
size of a typical dislocation pair in the Wigner crystal. $\xi_{KT}$
increases very sharply as the melting transition is approached and
diverges at the transition. Therefore it must become
larger than the decreasing $\xi_T$ before the melting transition
occurs, thus invalidating the Fukuyama-Lee-Rice
description which considers only phase-fluctuations \cite{FLRMod}.
We expect that as $T$ increases and dislocation pairs proliferate and
increase their size, the depinning transition will become rounded and a
more appropriate physical picture of the system at $E > 0$
would involve a combination of pinned and moving regions.
The effects of dislocations on the nonlinear conductivity have been studied
in numerical simulations by
Shi and Berlinsky for vortex lattices in type-II superconductors
at zero temperature \cite{Vlattice}.

Finally, the issue of the melting of the Wigner crystal
is also
straightforward. The melting temperature may be roughly
estimated from the Kosterlitz-Thouless theory and the
theoretical shear modulus \cite{KT}.
It is found to be of order $1K$ by Fisher \cite{KT}, much higher
than the temperature at which the observed $E_{Th}$ vanishes.
It is in fact much higher than the temperatures at which
these nonlinear transport measurements were
typically carried out \cite{Jiang}. Combining this with the
observations that we made in the last two paragraphs, we conclude that
it is unlikely that the vanishing of $E_{Th}$ observed in the
experiments \cite{I,II,III,Jiang} was due to the Wigner crystal
melting transition.

\section{Summary of Recent Experiments and Comparison to Theory}

So far we have focused on the theoretical aspects of the
problem of a sliding Wigner crystal in a strong magnetic field
when the sliding velocity is large.
In what follows, we give a brief review of the
relationship of our theoretical results to the
nonlinear conductivity measurements
reported in the literature.
Seemingly similar experiments obtain apparently inconsistent results.
We discuss these apparent inconsistencies in detail.
Experiments regarding the Hall resistivity were discussed earlier
in this paper; so here we will focus on the longitudinal
conductivity only.
We review and compare the DC transport experiments in
Sec.~IXA, and we compare the theory and the experiments in
Sec.~IXB.
AC + DC interference effects are briefly discussed in Sec.~IXC.

\subsection{Nonlinear DC Transport Experiments}

The basic idea of these experiments
is quite simple \cite{I,II,III,Jiang,CDWRev,FLRMod}:
a large enough DC electric field
will dislodge the pinned Wigner crystal and cause it to
slide. Therefore, by measuring the
conductivity as a function of the applied DC field,
one expects to observe a sudden increase of conductivity,
and the accompanying noise generated by the disorder
scattering off the sliding Wigner crystal.

There are to our knowledge four sets of published results
probing the nonlinear conductivity of the insulating phases
around the filling factor
$\nu = 1/5$ fractional quantum Hall state:
Goldman {\it et al.},
Phys. Rev. Lett. {\bf 65}, 2189 (1990)
(Ref.~\cite{I});
Williams {\it et al.}, Phys. Rev. Lett. {\bf 66}, 3285 (1991)
(Ref.~\cite{II});
Li {\it et al.}, Phys. Rev. Lett. {\bf 67}, 1630 (1991)
(Ref.~\cite{III});
and, Jiang {\it et al.}, Phys. Rev. B {\bf 44}, 8107 (1991)
(Ref.~\cite{Jiang}).
Among them, work by Jiang {\it et al.} \cite{Jiang}
is done at somewhat higher temperatures and covers a broader range
of applied field.
In the first half of this subsection, we shall discuss and
compare the first three experiments; in the second half,
we summarize the
findings from the last one.

\vskip 2.0truecm
\noindent
{\bf A1}

\vskip 1.0truecm
\noindent
A.~Sample quality

The mobility of the samples used in Ref.~\cite{I} was not given in
the paper.
However, a rising resistivity at
$\nu = 1/5$ as the temperature is
lowered is observed, although at temperatures reported
in the paper the
resistivity at $\nu = 1/5$ is a well-defined local
minimum.
The lowest temperature
state at
$\nu = 1/5$ is therefore insulating in these samples.
What impact, if any, it may have on the insulating phases
around $\nu = 1/5$ is not clear.

The mobilities quoted in Ref.~\cite{II} for various samples used
range from $4\times 10^6$ to $9\times 10^6~cm^2V^{-1}s^{-1}$, and those
quoted in Ref.~\cite{III} were somewhat lower $\sim 1\times 10^6$
Hall liquid state was observed. The size of the FQHE gap
at $\nu = 1/5$ was reported to be
$\sim 0.1K$ in Ref.~\cite{III}; it was not given
in Ref.~\cite{II}.

\vskip 1.0truecm
\noindent
B.~Threshold field, broad band noise, and their temperature
dependence

The observed apparent depinning threshold field is $\sim 1 mV cm^{-1}$
in Ref.~\cite{I}, which is similar to that in Ref.~\cite{III}, found
to be $\sim 1.2 mV cm^{-1}$, but differs greatly from the much
larger value of
$\sim 250 mV cm^{-1}$ found
in Ref.~\cite{II}.
It is certainly worth checking if in Ref.~\cite{II}
there is another threshold at fields similar to those in Refs.~\cite{I,III}
which may have escaped observation.
We also note here that the conventional four-terminal measurement was
done in Ref.~\cite{I}, but a two-terminal geometry was adopted in the
nonlinear conductivity measurement in Ref.~\cite{II}. In
Ref.~\cite{III}, both the four-terminal and
the two-terminal measurements were carried out in the low field (less than
$10 mV cm^{-1}$) regime, and were found to give essentially the same
results for the depinning threshold field and the nonlinear conductivity.

In addition, at fields above
the depinning threshold,
broad band noise (BBN) in the frequency range of $\sim 1KHz$
was observed and its power spectrum
measured in Refs.~\cite{I,III}. But in Ref.~\cite{I},
the external field at which the BBN was observed is greater by
a factor of $\sim 70$ than that at which the nonlinear
conductivity set in. No explanation was given for this rather
large discrepancy \cite{I}. In Ref.~\cite{III}, the threshold field
for the onset of BBN and that for the nonlinear conductivity were
found to agree with each other.

Despite the wide range of the observed depinning threshold field itself,
its temperature dependence was remarkably similar in Refs.~\cite{I,II,III}.
It decreases smoothly and continuously as temperature rises, and
becomes unobservable at about $\sim 120 mK$. It was suggested in
Refs.~\cite{I,II} that the disappearance of an apparent depinning field
could be due to the melting of the Wigner crystal at this temperature.
The BBN in Refs.~\cite{I,III}
had roughly the same temperature dependence as the depinning threshold field.
In particular,
it disappears at about the same temperature as the threshold field in
Ref.~\cite{I}, but in Ref.~\cite{III}, it was found the disappear earlier,
at $\sim 60 mK$. It was therefore cautioned in Ref.~\cite{III} that
it remains unclear what is the significance of these temperatures.

We should also note that although Refs.~\cite{I} and \cite{III} reported
similar threshold fields, the details of the depinning transition appear
to be rather different: In Ref.~\cite{III}, a step-function-like increase in
differential conductivity was
observed at the depinning transition while a gradual
and smooth one was found in Ref.~\cite{I} at and
above the depinning field.
The transition was the sharpest for the insulating phase
at $\nu=0.214$, slightly above $\nu = 1/5$ in Ref.~\cite{III}.
In Ref.~\cite{II} with much
larger depinning fields, the transition also appears very sharp for all
filling factors reported, down to $\nu \sim 0.15$.

\vskip 1.0truecm
\noindent
C.~Magnetic field dependence of the threshold field

In Refs.~\cite{II} and \cite{III}, studies were carried out for
the magnetic field, or the filling factor dependence of the depinning
threshold field at temperatures low enough so that the threshold field
does not vary with temperature. This information is useful as such dependences
near the transition from the insulating phase into the FQHE liquid
phase at and immediately near $\nu = 1/5$
may yield information about the nature of the phase
transition \cite{NLM}.

In Ref.~\cite{II}, the threshold field was found to decrease
rapidly and monotonically approaching the FQHE liquid phase at $\nu = 1/5$
on both sides of the magnetic field.
This differs from the observations in Ref.~\cite{III} in which
the threshold field was found to increase rapidly on both
sides of $\nu = 1/5$
as the (reentrant) insulating-FQHE transition was approached.
For large enough field, or small enough filling factor, as the
insulator-FQHE transition was far enough, the depinning threshold field
was found to increase
with the magnetic field in Ref.~\cite{III} as was in Ref.~\cite{II}.

\vskip 1.0truecm
\noindent
D.~Temperature dependence of the differential resistivity
above the apparent threshold field

In Ref.~\cite{III}, only the experiments done at the lowest
temperatures ($\sim 22 mK$) were reported, and it was found that
the differential resistivity is $T$-independent above
the threshold field at this
temperature.
In Ref.~\cite{II} with a much
larger depinning threshold field,
a wider temperature range (between $20 mK$ and $100 mK$) was examined.
It was found also
that the differential resistivity is $T$-independent above
the depinning transition, but strongly $T$-dependent
below it.
{}From these results, we may conclude that
above the threshold field,
conduction is no longer activated. We may also conclude
that
the main damping mechanism above
the threshold field should be the same for all temperatures in \cite{II}.

In Ref.~\cite{I}, the differential resistivity remains field-dependent
above the depinning transition and the temperature dependence is
difficult to establish clearly with the data given there.

\vskip 1.0truecm
\noindent
E.~Size of the insulating gap

Below the depinning threshold field, the conduction in the insulating
phase is presumably due to thermally excited charged defects/excitations.
If there is only a single mechanism for the thermal excitations,
the measurement of the temperature dependence of the conductivity
yields the size of the gap for such excitations.

The size of the insulating gap was not reported in Refs.~\cite{II,III}.
In Ref.~\cite{I},
the temperature dependence of the longitudinal
resistivity in the insulating phases
shows a complicated
behavior, apparently involving two regimes with different
temperature dependence in the temperature
range of 30 --- 100 mK.
The
insulating transport gap was reported
to be $\sim 0.5K$. We note that this gap is much larger than
the critical temperature ($\sim 0.12 K$) at which the threshold field
disappears, postulated to be related to the melting of the
Wigner crystal \cite{I}.
Furthermore, the
insulating gap is well-defined at temperatures above the so-called
melting temperature. This may be taken as a sign that the amplitude
of the charge density does not vary too much
in this temperature range,
since the size of the insulating gap is presumably
strongly dependent on the charge-density wave amplitude.

\vskip 2.0truecm
\noindent
{\bf A2} Summary of findings from Ref.~\cite{Jiang}
\vskip 1.0truecm

We believe that the samples used in \cite{Jiang} are probably of higher
quality than those used in Refs.~\cite{I,II,III}, because
the FQHE gap at $\nu = 1/5$ in the samples of Ref.~\cite{Jiang}
was $1.1K$, much larger than that in the others. Interestingly,
the activation gap in the insulating phases is larger by a factor of
3 --- 4, than that in \cite{I} where the $\nu =1/5$ state
is insulating at the lowest temperatures, for the same filling factors.
If the size of the insulating gap can be used as a measure of the
strength of the insulating phases, we then must conclude
that as sample quality improves, or as disorder becomes weaker, the
insulating phases become stronger. We believe this is strong
evidence supporting the notion that these insulating phases are
primarily due to interaction, and not disorder.

Threshold nonlinearity in conductivity was also observed in \cite{Jiang}
for applied voltages $\sim$ 0.1 mV across samples of
typical dimensions $3mm \times 3mm$, giving a
depinning field $\sim 1.3 mV cm^{-1}$, similar to those in
\cite{I} and \cite{III}.
However, at and above the threshold field, there is no
clear step-wise change in the resistivity. This is similar
to \cite{I}, but differs from \cite{III} in which lower
temperatures were reached.

Although the threshold field diminishes
at a temperature about 100 mK,
below and well above
the threshold field (within a factor of 10 of $E_{Th}$),
$\rho_{xx}$ remains thermally activated.
This differs qualitatively from the observations in \cite{II}
which are done at generally lower temperatures.
Furthermore, the conduction activation gap remains the same
below and well above the threshold field. In addition, not only
the gap itself is an order of magnitude larger than the
so-called critical temperature, but also the activated behavior extends
above the critical temperature by
at least a factor of 3 --- 4.
Two related
conclusions that one can draw from this important observation are:
1). The dominant damping mechanism giving rise to
resistivity above and below the
low-field threshold field
ought to be the same; and 2). The solid retains its identity up
to temperatures higher than the critical temperature by at least a
factor of 3 -- 4.
The first point
has a natural explanation
from the coupled motion of a sliding Wigner
crystal with thermally activated carriers.
The second point supports the notion that the critical $T$ at which
the apparent threshold disappears does not signify the melting
temperature of the Wigner crystal.
We offer a speculation for this experimental observation in the next
subsection.

A voltage breakdown is observed in the high field limit
around $50$ mV in \cite{Jiang}, in the same range as
where the work in \cite{II} was carried out.
It is suggested by the authors of
\cite{Jiang} that the power dissipation observed in
\cite{Jiang} is consistent with an excessive electron heating
interpretaion.

In addition, it is also found in \cite{Jiang} that
the activation gap in the insulating phases
may change upon thermal cycling to room temperature by
a factor of two, as does the conductivity threshold.
Upon thermal cycling to room temperature,
the disorder configuration must change due to diffusion or other
thermally activated processes.
This suggests that disorder also plays an important
role in determining the insulating gap.
There is also a correlation between the density and
the activation gap, which may be taken as evidence
for the role of interaction.
The insulating gap decreases toward $\nu = 1/5$ in its
vicinity, similar to the work in \cite{I}.
Although the activation gap itself
changes upon thermal cycling,
its variation with filling factor
is the same within a cycle. Therefore the filling-factor
dependence of the insulating gap around the
$\nu = 1/5$ fractional quantum Hall liquid state
may be intrinsic to a
clean system.

We see from the above discussion that the experimental
situation is not very clear, and different experiments often
contradict each other in important details. In the next subsection,
we attempt to interprete some of these
experiments related to the nonlinear conductivity
in light of our theoretical results.

\subsection{Comparison of Theory and Experiments in the DC limit}

According
to the behavior of the conductivity at fields above threshold,
the current experiments reviewed above fall into two categories:

\noindent
1). The {\it differential} conductivity is
voltage- and temperature-independent above threshold.
This is reported in Ref.~\cite{II}
which finds
a very large threshold field ($\sim 50 mV$) at the
lowest temperatures $\sim 36 mK$, and in Ref.~\cite{III}
which finds a much smaller threshold field ($\sim 0.2 mV$) and is
done at very low temperatures
$\sim~20mK$.

\noindent
2). The differential conductivity above the threshold field
continues to depend on
voltage and temperature,
as reported in Ref.~\cite{I}
and Ref.~\cite{Jiang}.
The
threshold field is found to be small ($\sim 0.2 mV$) \cite{I,Jiang}.
The temperature dependence of the differential
resistivity is
found in Ref.~\cite{Jiang}
to be thermally
activated with an activation gap identical to that
below threshold,
for temperatures between $\sim~80~mK$ and $\sim~200~mK$.
However, it is also reported in
Ref.~\cite{Jiang} that there exists
a second threshold field at $\sim 30 mV$ above which the
differential resistivity is no longer field-
or temperature-dependent, and its value is consistent with those
in Ref.~\cite{II,III}.

We first compare our theory with
experiments done at the lowest temperatures where
thermally activated free carriers are unimportant,
{\it i.e.\/}, the experiments in the first category described above.
The predicted $I-V$ curve from Eq.~\ref{eandj1} agrees
very well with the low temperature
experiments (Compare Figs.~1 and 2 from this work
to Fig.~2 in Ref.~\cite{III} and
Fig.~2 in Ref.~\cite{II}) -- that is, the differential
DC resistivity does not depend on the driving field, or equivalently
on the resulting current above the sliding threshold.
In addition,
the intercept of the $I-V$ curve from the high velocity perturbation
theory gives also a measure of the sliding threshold field.
{}From Eqs.~\ref{eandj1} and \ref{jxande}, one sees that
this statement can only be true
when, in plotting
the $I-V$, the field $V$ is measured
along the direction of the current flow.

We now consider the large field $( > 30 mV)$, therefore
high velocity, data reported in Ref.~\cite{Jiang} in which
electron heating was invoked to explain the data.
We wish to propose an alternative possibility.
The differential conductivity in the
$ V > 30 mV$ regime is field and temperature
independent, and has roughly the same value
($\sim 10^5 Ohms$) as the lowest temperature data
in Ref.~\cite{III}.
This experimental fact \cite{footnote2}
is consistent with our
result that the large-velocity  differential resistivity is
$\sigma^F_{xx}$-independent. We propose that the second
apparent threshold field in fact signifies a crossover from
a regime where disorder effects enter non-perturbatively
to one where they can be treated perturbatively as was done
in this paper.
As the velocity decreases, it is expected
that our second-order perturbation theory becomes no longer valid, and that
$\sigma^F_{xx}$ affects the differential conductivity,
In this limit, the
effective damping constant may be of order
$\lambda_1 \propto (\sigma^F_{xx})^{-1}$, and would lead to an activated
temperature dependence as observed. This is, however, beyond the range
we can treat by the present perturbation theory.

Our predicted $V-I$
characteristic along the direction
of the current $\vec j$ for a sliding
Wigner crystal: $E = \rho_m \lambda j / \rho_c^2 + E_{Th}$
agrees very well with the experiments in III done
at the lowest temperatures and the high-field results
in II and \cite{Jiang}. The details of the pinning-depinning
transition are not accessible from the present
perturbation theory and there is no clear picture from
the experiments. The strong temperature dependence of the
differential conductivity above the small threshold field
observed in \cite{Jiang} at somewhat elevated temperatures
indicates the possible role of thermally excited free carriers in the
sliding regime beyond that revealed by our perturbation theory.

We finally speculate on the possible cause of the disappearence of the
apprarent threshold field at temperatures around $100 - 200~mK$
observed in several recent experiments \cite{I,II,III,Jiang}.
It was suggested that it may signify the melting of the Wigner crystal in
Refs.~\cite{I,II}, but this suggestion was called into
question in Refs.~\cite{III,Jiang}. We here wish to echo the
cautions that have already been raised. First of all, one could
estimate roughly the classical melting temperature from the shear
modulus, assuming a Kosterlitz-Thouless (KT) melting mechanism \cite{KT}.
We find a temperature $T_{KT}$ around $400 - 600~mK$. Secondly,
the magneto-optical measurements \cite{MOptics} indicate
the possibility of a still higher melting temperature of $1 - 2~K$.
Thirdly, above the so-determined ``melting temperature'', the measured
$I-V$ curves continue to be highly nonlinear \cite{I,Jiang}, which
is also inconsistent with the notion that the Wigner crystal
has already melted into
a liquid.
A natural alternative to this interpretation is to assume what has
been observed is in fact the thermal depinning of the Wigner crystal.
At temperatures above $100 - 200~mK$ in these
experiments, we propose that
thermal fluctuations destroy the pinning due to disorder
potential, and the subsequent sliding motion is one of a solid with
large thermal fluctuations and subjected to disorder.

\subsection{Theory and Experiments for AC + DC Interferences}

Finally, we briefly remark on some recent experimental work
\cite{LiTsui1,LiTsui2} on
the AC + DC interference effects in a sliding Wigner crystal
around filling factor $\nu = 1/5$.
In the absence of a DC current, the AC experiments by Li {\it et al.\/},
in
the frequency range of $\sim ~30 - 100$ $MHz$
have established the capacitive response of the pinned Wigner crystal
\cite{LiTsui1}. A detailed analysis and the possible implications for
the nature of the insulating phases from the pure AC measurements were
given in their original paper \cite{LiTsui1}.

So far, there have been no reports on the observation of the
Shapiro anomaly in the DC response due to an AC field.
However, preliminary data have established the interference effects in
the AC response in the presence of a large DC current.
This situation is not surprising since from the experience
with the CDW systems \cite{CDWRev}, the requirements on the sample
quality are usually less stringent for the observation of the latter effect.

In Fig.~3, we give a qualitative sketch of the calculated inductive
anomaly from Eqs.~\ref{acresponse2}
and \ref{acresponse4} in Sec.~VIB. The dotted portion of the
curve indicates that our linear AC response theory is not reliable
at AC frequencies very close to the
washboard frequency of the DC current.
Nonetheless, its overall shape is in good agreement with the experimental
observations, where it was found
that the capacitive AC response in the absence of the DC current
becomes inductive in its presence at frequencies below $2~MHz$ \cite{LiTsui2}.
However, a direct transition from the
inductive response to capacitive response as the AC frequency increases
has not been observed due to the limited AC
frequency range
attainable in the interference effect measurements \cite{LiTsui2}.

Our theory (see Eq.~\ref{acresponse2} and
Eq.~\ref{acresponse6})
also suggests that at low $AC$ frequencies, the linear
slope of $Im \rho(\omega)$ in $\omega$ at small $\omega$ would depend on
the DC current as $\sim 1/j_{DC}^2$.
To observe this effect experimentally, care must be taken to
ensure that the $AC$-response remains in the linear
regime, {\it i.e.\/}, the $AC$-component amplitude must be reduced
along with its frequency $\omega$ to ensure the
accuracy of the small parameter expansion
of the Bessel's function (see Eq.~\ref{acdc} and Eq.~\ref{acresponse}).

Clearly, further experiments are needed to test other aspects
of our theory in terms of the AC + DC interference effects. We refer
the reader to Sec.~VI for a more detailed description of what
one might expect from the present perturbation theory.

\section{Conclusions}

In conclusion, we have studied the sliding
motion of a Wigner crystal in a strong magnetic field.
We obtain the form of the nonlinear resistivity  in
the regime of large sliding velocity. The Hall resistance of a
sliding Wigner crystal is found not to be changed by
disorder scattering. We predict the AC + DC interference
effects and compare the present case to the CDW systems
and point out the differences. We also give a brief summary
of the available experiments that measure the nonlinear
conductivity of the insulating phases around filling
factor $\nu = 1/5$. At low temperatures and/or large
depinning fields, the experimentally observed
nonlinear $I-V$ curves are in
agreement with our theoretical results. In the regime of
smaller
fields and elevated temperatures, results from various
experiments
differ in important ways as we discussed in the paper,
and a definitive comparison to the present theory is
difficult to make.

\acknowledgements

We thank Drs. Y. P. Li, D. C. Tsui, H. W. Jiang, and H. Stormer
for discussions of the current experiments. We are particularly
grateful to Y. P. Li and D. C. Tsui for showing us their
data prior to publication and we wish to thank Dr. S. N. Coppersmith for
many helpful conversations and a careful reading of the manuscript.

\newpage
\centerline{Figure Captions}


\vskip 1.0truecm

\noindent
Fig.~1. Theoretical nonlinear $I-V$ properties of a
sliding Wigner crystal in a strong magnetic field in the
presence of disorder for the Hall bar geometry.
$I$ is the total current, $V_L$ is the field component
parallel to $I$, and $V_H$ is the field component
perpendicular to $I$. The crossing point of the two curves is where the
Hall angle becomes $45^o$.
The dotted portion in $V_L$ indicates
that the present perturbation theory is not
reliable when the external field is only slightly greater
than the depinning threshold field, indicated
by $V_{Th}$. We have assumed $\omega_c$ greater than $\lambda$.
The dashed portion is the conductivity due to free carriers
at fields below the threshold field. It is extended to fields
slightly above $V_{Th}$ for clarity of illustration.
The line of $V_H$ as a function of $I$ is valid and
extends all the way to zero $I$ provided that the transport
current is due to the sliding of the whole Wigner crystal
under a combined total external field $\sqrt{V_H^2+V_L^2}$ greater
than $V_{Th}$.

\vskip 1.0truecm

\noindent
Fig.~2. The theoretical differential resistivity
of a sliding Wigner crystal in a strong magnetic field as a function
of the external field for the Hall bar geometry.
The depinning threshold field is indicated
by $V_{Th}$. We have assumed $\omega_c$ greater than $\lambda$.
The dotted portion of the $dV_L/dI$ curve indicates that
the theory is not reliable near threshold field.
The dashed portion of the
$dV_L/dI$ curve
indicates the conduction of thermally excited free carriers
in this regime where $dV_L/dI = \rho^F_{xx}$, the
diagonal resistivity of the free carriers.

\vskip 1.0truecm

\noindent
Fig.~3. The
inductive anomaly in the imaginary part
of the AC response of a
Wigner crystal in the presence of a large DC current.
The dotted portion indicates that the linear AC response theory
is not reliable near frequency locking.


\begin{references}
\bibitem{footnote1} We should also caution that
$\sigma_{xx}^F$ is itself $(\vec k,\omega)$-dependent.
In our discussion, we have implicitly assumed that the
$(\vec k,\omega)$-range of interest in the sliding
Wigner crystal lies within the conductivity limit
of $\sigma_{xx}^F$. For more discussion, see Appendix A
in Ref.~\cite{NLM}.
\bibitem{footnote2} This
is despite the fact that their behaviors below
the upper threshold field are very different, and dependent on
temperature \cite{II,III,Jiang}, often in an activated
fashion suggesting the involvement
of dissipation in the screened longitudinal channel \cite{Jiang}.
\end{references}
\end{document}